\documentclass[twocolumn,resetfootnote]{aastex701}

\usepackage{soul}
\usepackage{color}
\usepackage{xspace}
\usepackage{graphicx}
\usepackage{amsmath}
\usepackage{amssymb}
\usepackage{yfonts}
\usepackage{xcolor}
\usepackage{float}
\hypersetup{pdfauthor={Name}}
\usepackage{stackengine}
\usepackage{xparse}
\usepackage{multirow}
\usepackage{longtable}
\usepackage{colortbl}
\usepackage{csquotes}
\usepackage{enumitem}
\usepackage[T1]{fontenc}
\usepackage{ae,aecompl}
\usepackage{rotating} 
\usepackage{tikz}
\usepackage{upgreek} 

\graphicspath{{./}{figures/}}

\shorttitle{GRB\,230307A}
\shortauthors{Dalessi et al.}

\begin{document}

\setlength{\LTcapwidth}{\textwidth}

\title{Fermi-GBM Observations of GRB 230307A: An Exceptionally Bright Long-Duration Gamma-ray Burst with an Associated Kilonova} 


\correspondingauthor{S. Dalessi}
\email{sd0104@uah.edu}



\def\shrinkage{2.1mu}
\def\vecsign{\mathchar"017E}
\def\dvecsign{\smash{\stackon[-1.95pt]{\mkern-\shrinkage\vecsign}{\rotatebox{180}{$\mkern-\shrinkage\vecsign$}}}}
\def\dblvec#1{\def\useanchorwidth{T}\stackon[-4.2pt]{#1}{\,\dvecsign}}
\stackMath
\def\Avec{\vec{A}}
\def\Bvec{\vec{B}}
\def\Cvec{\vec{C}}
\def\Dvec{\vec{D}}
\def\Evec{\vec{E}}
\def\Fvec{\vec{F}}
\def\Gvec{\vec{G}}
\def\Hvec{\vec{H}}
\def\Ivec{\vec{I}}
\def\Jvec{\vec{J}}
\def\Kvec{\vec{K}}
\def\Lvec{\vec{L}}
\def\Mvec{\vec{M}}
\def\Nvec{\vec{N}}
\def\Ovec{\vec{O}}
\def\Pvec{\vec{P}}
\def\Qvec{\vec{Q}}
\def\Rvec{\vec{R}}
\def\Svec{\vec{S}}
\def\Tvec{\vec{T}}
\def\Uvec{\vec{U}}
\def\Vvec{\vec{V}}
\def\Wvec{\vec{W}}
\def\Xvec{\vec{X}}
\def\Yvec{\vec{Y}}
\def\Zvec{\vec{Z}}
\def\Ahat{\hat{A}}
\def\Bhat{\hat{B}}
\def\Chat{\hat{C}}
\def\Dhat{\hat{D}}
\def\Ehat{\hat{E}}
\def\Fhat{\hat{F}}
\def\Ghat{\hat{G}}
\def\Hhat{\hat{H}}
\def\Ihat{\hat{I}}
\def\Jhat{\hat{J}}
\def\Khat{\hat{K}}
\def\Lhat{\hat{L}}
\def\Mhat{\hat{M}}
\def\Nhat{\hat{N}}
\def\Ohat{\hat{O}}
\def\Phat{\hat{P}}
\def\Qhat{\hat{Q}}
\def\Rhat{\hat{R}}
\def\Shat{\hat{S}}
\def\That{\hat{T}}
\def\Uhat{\hat{U}}
\def\Vhat{\hat{V}}
\def\What{\hat{W}}
\def\Xhat{\hat{X}}
\def\Yhat{\hat{Y}}
\def\Zhat{\hat{Z}}
\def\Aten{\dblvec{A}}
\def\Bten{\dblvec{B}}
\def\Cten{\dblvec{C}}
\def\Dten{\dblvec{D}}
\def\Eten{\dblvec{E}}
\def\Ften{\dblvec{F}}
\def\Gten{\dblvec{G}}
\def\Hten{\dblvec{H}}
\def\Iten{\dblvec{I}}
\def\Jten{\dblvec{J}}
\def\Kten{\dblvec{K}}
\def\Lten{\dblvec{L}}
\def\Mten{\dblvec{M}}
\def\Nten{\dblvec{N}}
\def\Oten{\dblvec{O}}
\def\Pten{\dblvec{P}}
\def\Qten{\dblvec{Q}}
\def\Rten{\dblvec{R}}
\def\Sten{\dblvec{S}}
\def\Tten{\dblvec{T}}
\def\Uten{\dblvec{U}}
\def\Vten{\dblvec{V}}
\def\Wten{\dblvec{W}}
\def\Xten{\dblvec{X}}
\def\Yten{\dblvec{Y}}
\def\Zten{\dblvec{Z}}
\def\Aol{$\mathbbm{A}$}
\def\Bol{$\mathbbm{B}$}
\def\Col{$\mathbbm{C}$}
\def\Dol{$\mathbbm{D}$}
\def\Eol{$\mathbbm{E}$}
\def\Fol{$\mathbbm{F}$}
\def\Gol{$\mathbbm{G}$}
\def\Hol{$\mathbbm{H}$}
\def\Iol{$\mathbbm{I}$}
\def\Jol{$\mathbbm{J}$}
\def\Kol{$\mathbbm{K}$}
\def\Lol{$\mathbbm{L}$}
\def\Mol{$\mathbbm{M}$}
\def\Nol{$\mathbbm{N}$}
\def\Ool{$\mathbbm{O}$}
\def\Pol{$\mathbbm{P}$}
\def\Qol{$\mathbbm{Q}$}
\def\Rol{$\mathbbm{R}$}
\def\Sol{$\mathbbm{S}$}
\def\Tol{$\mathbbm{T}$}
\def\Uol{$\mathbbm{U}$}
\def\Vol{$\mathbbm{V}$}
\def\Wol{$\mathbbm{W}$}
\def\Xol{$\mathbbm{X}$}
\def\Yol{$\mathbbm{Y}$}
\def\Zol{$\mathbbm{Z}$}
\def\avec{\vec{a}}
\def\bvec{\vec{b}}
\def\cvec{\vec{c}}
\def\dvec{\vec{d}}
\def\evec{\vec{e}}
\def\fvec{\vec{f}}
\def\gvec{\vec{g}}
\def\hvec{\vec{h}}
\def\ivec{\vec{i}}
\def\jvec{\vec{j}}
\def\kvec{\vec{k}}
\def\lvec{\vec{l}}
\def\mvec{\vec{m}}
\def\nvec{\vec{n}}
\def\ovec{\vec{o}}
\def\pvec{\vec{p}}
\def\qvec{\vec{q}}
\def\rvec{\vec{r}}
\def\svec{\vec{s}}
\def\tvec{\vec{t}}
\def\uvec{\vec{u}}
\def\vvec{\vec{v}}
\def\wvec{\vec{w}}
\def\xvec{\vec{x}}
\def\yvec{\vec{y}}
\def\zvec{\vec{z}}
\def\ahat{\hat{a}}
\def\bhat{\hat{b}}
\def\chat{\hat{c}}
\def\dhat{\hat{d}}
\def\ehat{\hat{e}}
\def\fhat{\hat{f}}
\def\ghat{\hat{g}}
\def\hhat{\hat{h}}
\def\ihat{\hat{i}}
\def\jhat{\hat{j}}
\def\khat{\hat{k}}
\def\lhat{\hat{l}}
\def\mhat{\hat{m}}
\def\nhat{\hat{n}}
\def\ohat{\hat{o}}
\def\phat{\hat{p}}
\def\qhat{\hat{q}}
\def\rhat{\hat{r}}
\def\shat{\hat{s}}
\def\that{\hat{t}}
\def\uhat{\hat{u}}
\def\vhat{\hat{v}}
\def\what{\hat{w}}
\def\xhat{\hat{x}}
\def\yhat{\hat{y}}
\def\zhat{\hat{z}}
\def\aten{\dblvec{a}}
\def\bten{\dblvec{b}}
\def\cten{\dblvec{c}}
\def\dten{\dblvec{d}}
\def\eten{\dblvec{e}}
\def\ften{\dblvec{f}}
\def\gten{\dblvec{g}}
\def\hten{\dblvec{h}}
\def\iten{\dblvec{i}}
\def\jten{\dblvec{j}}
\def\kten{\dblvec{k}}
\def\lten{\dblvec{l}}
\def\mten{\dblvec{m}}
\def\nten{\dblvec{n}}
\def\oten{\dblvec{o}}
\def\pten{\dblvec{p}}
\def\qten{\dblvec{q}}
\def\rten{\dblvec{r}}
\def\sten{\dblvec{s}}
\def\tten{\dblvec{t}}
\def\uten{\dblvec{u}}
\def\vten{\dblvec{v}}
\def\wten{\dblvec{w}}
\def\xten{\dblvec{x}}
\def\yten{\dblvec{y}}
\def\zten{\dblvec{z}}
\def\aol{$\mathbbm{a}$}
\def\bol{$\mathbbm{b}$}
\def\col{$\mathbbm{c}$}
\def\dol{$\mathbbm{d}$}
\def\eol{$\mathbbm{e}$}
\def\fol{$\mathbbm{f}$}
\def\gol{$\mathbbm{g}$}
\def\hol{$\mathbbm{h}$}
\def\iol{$\mathbbm{i}$}
\def\jol{$\mathbbm{j}$}
\def\kol{$\mathbbm{k}$}
\def\lol{$\mathbbm{l}$}
\def\mol{$\mathbbm{m}$}
\def\nol{$\mathbbm{n}$}
\def\ool{$\mathbbm{o}$}
\def\pol{$\mathbbm{p}$}
\def\qol{$\mathbbm{q}$}
\def\rol{$\mathbbm{r}$}
\def\sol{$\mathbbm{s}$}
\def\tol{$\mathbbm{t}$}
\def\uol{$\mathbbm{u}$}
\def\vol{$\mathbbm{v}$}
\def\wol{$\mathbbm{w}$}
\def\xol{$\mathbbm{x}$}
\def\yol{$\mathbbm{y}$}
\def\zol{$\mathbbm{z}$}
\newcommand{\eps}{\epsilon}
\newcommand{\veps}{\varepsilon}
\newcommand{\vtheta}{\vartheta}
\newcommand{\vphi}{\varphi}
\newcommand{\vrho}{\varrho}
\def\alphavec{\vec{\alpha}}
\def\nuvec{\vec{\nu}}
\def\betavec{\vec{\beta}}
\def\xivec{\vec{\xi}}
\def\Xivec{\vec{\Xi}}
\def\gammavec{\vec{\gamma}} 
\def\Gammavec{\vec{\Gamma}}
\def\deltavec{\vec{\delta}} 
\def\Deltavec{\vec{\Delta}}
\def\pivec{\vec{\pi}} 
\def\Pivec{\vec{\Pi}}
\def\epsvec{\vec{\eps}} 
\def\vepsvec{\vec{\veps}} 
\def\rhovec{\vec{\rho}}
\def\vrhovec{\vec{\vrho}}
\def\zetavec{\vec{\zeta}}
\def\sigmavec{\vec{\sigma}}
\def\Sigmavec{\vec{\Sigma}}
\def\etavec{\vec{\eta}}
\def\tauvec{\vec{\tau}}
\def\thetavec{\vec{\theta}}
\def\vthetavec{\vec{\vtheta}}
\def\Thetavec{\vec{\Theta}}
\def\upsilonvec{\vec{\upsilon}}
\def\Upsilonvec{\vec{\Upsilon}}
\def\iotavec{\vec{\iota}}
\def\phivec{\vec{\phi}}
\def\vphivec{\vec{\vphi}}
\def\Phivec{\vec{\Phi}}
\def\kappavec{\vec{\kappa}}
\def\chivec{\vec{\chi}}
\def\lambdavec{\vec{\lambda}}
\def\Lambdavec{\vec{\Lambda}}
\def\psivec{\vec{\psi}}
\def\Psivec{\vec{\Psi}}
\def\muvec{\vec{\mu}}
\def\omegavec{\vec{\omega}}
\def\Omegavec{\vec{\Omega}}
\def\alphahat{\hat{\alpha}}
\def\nuhat{\hat{\nu}}
\def\betahat{\hat{\beta}}
\def\xihat{\hat{\xi}}
\def\Xihat{\hat{\Xi}}
\def\gammahat{\hat{\gamma}} 
\def\Gammahat{\hat{\Gamma}}
\def\deltahat{\hat{\delta}} 
\def\Deltahat{\hat{\Delta}}
\def\pihat{\hat{\pi}} 
\def\Pihat{\hat{\Pi}}
\def\epshat{\hat{\eps}} 
\def\vepshat{\hat{\veps}} 
\def\rhohat{\hat{\rho}}
\def\vrhohat{\hat{\vrho}}
\def\zetahat{\hat{\zeta}}
\def\sigmahat{\hat{\sigma}}
\def\Sigmahat{\hat{\Sigma}}
\def\etahat{\hat{\eta}}
\def\tauhat{\hat{\tau}}
\def\thetahat{\hat{\theta}}
\def\vthetahat{\hat{\vtheta}}
\def\Thetahat{\hat{\Theta}}
\def\upsilonhat{\hat{\upsilon}}
\def\Upsilonhat{\hat{\Upsilon}}
\def\iotahat{\hat{\iota}}
\def\phihat{\hat{\phi}}
\def\vphihat{\hat{\vphi}}
\def\Phihat{\hat{\Phi}}
\def\kappahat{\hat{\kappa}}
\def\chihat{\hat{\chi}}
\def\lambdahat{\hat{\lambda}}
\def\Lambdahat{\hat{\Lambda}}
\def\psihat{\hat{\psi}}
\def\Psihat{\hat{\Psi}}
\def\muhat{\hat{\mu}}
\def\omegahat{\hat{\omega}}
\def\Omegahat{\hat{\Omega}}
\def\alphaten{\dblvec{\alpha}}
\def\nuten{\dblvec{\nu}}
\def\betaten{\dblvec{\beta}}
\def\xiten{\dblvec{\xi}}
\def\Xiten{\dblvec{\Xi}}
\def\gammaten{\dblvec{\gamma}} 
\def\Gammaten{\dblvec{\Gamma}}
\def\deltaten{\dblvec{\delta}} 
\def\Deltaten{\dblvec{\Delta}}
\def\piten{\dblvec{\pi}} 
\def\Piten{\dblvec{\Pi}}
\def\epsten{\dblvec{\eps}} 
\def\vepsten{\dblvec{\veps}} 
\def\rhoten{\dblvec{\rho}}
\def\vrhoten{\dblvec{\vrho}}
\def\zetaten{\dblvec{\zeta}}
\def\sigmaten{\dblvec{\sigma}}
\def\Sigmaten{\dblvec{\Sigma}}
\def\etaten{\dblvec{\eta}}
\def\tauten{\dblvec{\tau}}
\def\thetaten{\dblvec{\theta}}
\def\vthetaten{\dblvec{\vtheta}}
\def\Thetaten{\dblvec{\Theta}}
\def\upsilonten{\dblvec{\upsilon}}
\def\Upsilonten{\dblvec{\Upsilon}}
\def\iotaten{\dblvec{\iota}}
\def\phiten{\dblvec{\phi}}
\def\vphiten{\dblvec{\vphi}}
\def\Phiten{\dblvec{\Phi}}
\def\kappaten{\dblvec{\kappa}}
\def\chiten{\dblvec{\chi}}
\def\lambdaten{\dblvec{\lambda}}
\def\Lambdaten{\dblvec{\Lambda}}
\def\psiten{\dblvec{\psi}}
\def\Psiten{\dblvec{\Psi}}
\def\muten{\dblvec{\mu}}
\def\omegaten{\dblvec{\omega}}
\def\Omegaten{\dblvec{\Omega}}
\def\alphaol{$\mathbb{\alpha}$}
\def\nuol{$\mathbb{\nu}$}
\def\betaol{$\mathbb{\beta}$}
\def\xiol{$\mathbb{\xi}$}
\def\Xiol{$\mathbb{\Xi}$}
\def\gammaol{$\mathbb{\gamma}$}
\def\Gammaol{$\mathbb{\Gamma}$}
\def\deltaol{$\mathbb{\delta}$}
\def\Deltaol{$\mathbb{\Delta}$}
\def\piol{$\mathbb{\pi}$}
\def\Piol{$\mathbb{\Pi}$}
\def\epsol{$\mathbb{\eps}$}
\def\vepsol{$\mathbb{\veps}$}
\def\rhool{$\mathbb{\rho}$}
\def\vrhool{$\mathbb{\vrho}$}
\def\zetaol{$\mathbb{\zeta}$}
\def\sigmaol{$\mathbb{\sigma}$}
\def\Sigmaol{$\mathbb{\Sigma}$}
\def\etaol{$\mathbb{\eta}$}
\def\tauol{$\mathbb{\tau}$}
\def\thetaol{$\mathbb{\theta}$}
\def\vthetaol{$\mathbb{\vtheta}$}
\def\Thetaol{$\mathbb{\Theta}$}
\def\upsilonol{$\mathbb{\upsilon}$}
\def\Upsilonol{$\mathbb{\Upsilon}$}
\def\iotaol{$\mathbb{\iota}$}
\def\phiol{$\mathbb{\phi}$}
\def\vphiol{$\mathbb{\vphi}$}
\def\Phiol{$\mathbb{\Phi}$}
\def\kappaol{$\mathbb{\kappa}$}
\def\chiol{$\mathbb{\chi}$}
\def\lambdaol{$\mathbb{\lambda}$}
\def\Lambdaol{$\mathbb{\Lambda}$}
\def\psiol{$\mathbb{\psi}$}
\def\Psiol{$\mathbb{\Psi}$}
\def\muol{$\mathbb{\mu}$}
\def\omegaol{$\mathbb{\omega}$}
\def\Omegaol{$\mathbb{\Omega}$}
\def\cross{\times}
\def\dot{\cdot}
\def\del{\nabla}
\def\delcross{\nabla \times}
\def\deldot{\nabla \cdot}
\def\delsq{\nabla^2}
\newcommand{\rarrow}{\Rightarrow}
\newcommand{\rrarrow}{\Longrightarrow}
\newcommand{\larrow}{\Leftarrow}
\newcommand{\llarrow}{\Longleftarrow}
\newcommand{\lrarrow}{\Leftrightarrow}
\newcommand{\llrrarrow}{\iff}
\newcommand{\nperp}{\not\perp}
\let\oldinf\inf
\renewcommand{\inf}{\infty}
\def\wbox{\square}
\def\bbox{\blacksquare}
\def\deg{^{\circ}}

\newcommand{\rough}[1]{{\color{brown} #1}}
\newcommand{\todo}[1]{\textbf{\color{red} Todo: #1}} 

\newcommand{\trm}{\textrm}
\newcommand{\tbf}{\textbf}
\newcommand{\tul}{\underline}
\newcommand{\tit}{\textit}
\newcommand{\texp}[1]{$^{\textrm{#1}}$}
\newcommand{\tqu}{\enquote}
\newcommand{\pref}[1]{(\pageref{#1})}
\newcommand{\eref}[1]{equation \eqref{#1}}
\newcommand{\avg}[1]{\overline{#1}}
\newcommand{\p}[1]{\left( #1 \right)}
\newcommand{\pp}[1]{\left[ #1 \right]}
\newcommand{\psqu}[1]{\left\{ #1 \right\}}
\newcommand{\pang}[1]{\left\langle #1 \right\rangle}
\newcommand{\abs}[1]{\left| #1 \right|}
\newcommand{\dabs}[1]{\left\lVert #1 \right\rVert}
\newcommand{\eval}[2]{\rvert_{#1}^{#2}}
\newcommand{\Eval}[2]{\Bigg\rvert_{#1}^{#2}}
\newcommand{\e}[1]{\times 10^{#1}}
\newcommand{\dv}[2]{\frac{d #1}{d #2}}
\newcommand{\ndv}[3]{\frac{d^{#1} #2}{d #3^{#1}}}
\newcommand{\pdv}[2]{\frac{\partial #1}{\partial #2}}
\newcommand{\npdv}[3]{\frac{\partial^{#1} #2}{\partial #3^{#1}}}
\newcommand{\ulabel}[2]{\underset{\mathclap{\substack{\uparrow\\#2}}}{#1}}
\newcommand{\llabel}[2]{\overset{\mathclap{\substack{#2\\\downarrow}}}{#1}}
\newcommand{\ublabel}[2]{\overbrace{#1}^{\mathclap{\substack{#2}}}}
\newcommand{\lblabel}[2]{\underbrace{#1}_{\mathclap{\substack{#2}}}}
\newcommand{\uslabel}[2]{\overbracket{#1}^{\mathclap{\substack{#2}}}}
\newcommand{\lslabel}[2]{\underbracket{#1}_{\mathclap{\substack{#2}}}}
\let\oldlim\lim
\renewcommand{\lim}[2]{\oldlim\limits_{{#1} \rightarrow {#2}}}
\let\oldsum\sum
\renewcommand{\sum}[2]{\oldsum\limits_{#1}^{#2}}
\let\oldprod\prod
\renewcommand{\prod}[2]{\oldprod\limits_{#1}^{#2}}
\let\oldint\int
\renewcommand{\int}[2]{\oldint\limits_{#1}^{#2}}
\newcommand{\dint}[4]{\oldint\limits_{#1}^{#2} \oldint\limits_{#3}^{#4}}
\newcommand{\tint}[6]{\oldint\limits_{#1}^{#2} \oldint\limits_{#3}^{#4} \oldint\limits_{#5}^{#6}}
\def\lint{\int_{l}}
\def\sint{\iint\limits_{S}}
\def\vint{\iiint\limits_{V}}
\def\olint{\oint\limits_{l}}
\def\osint{\oiint\limits_{S}}
\def\ovint{\oiint\limits_{V}\hspace{-10.9pt} \oldint_{}^{}}
\newcommand{\eq}[1]{\begin{equation*} #1 \end{equation*}}
\newcommand{\eql}[2]{\begin{equation} \label{#1} #2 \end{equation}}
\newcommand{\teq}[1]{$ #1 $}
\let\oldmatrix\matrix
\renewcommand{\matrix}[1]{$\begin{pmatrix} #1 \end{pmatrix}$}


\def\nm{\mbox{~nm}} 
\def\mum{\mbox{~\mu\hbox{m}}} 
\def\mm{\mbox{~mm}} 
\def\cm{\mbox{~cm}} 
\def\m{\mbox{~m}} 
\def\km{\mbox{~km}} 
\def\pc{\mbox{~pc}} 
\def\kpc{\mbox{~kpc}} 
\def\Mpc{\mbox{~Mpc}} 
\def\Gpc{\mbox{~Gpc}} 
\def\erg{\mbox{~erg}}
\def\eV{\mbox{~eV}} 
\def\keV{\mbox{~keV}} 
\def\MeV{\mbox{~MeV}} 
\def\GeV{\mbox{~GeV}} 
\def\TeV{\mbox{~TeV}} 
\def\Hz{\mbox{~Hz}} 
\def\kHz{\mbox{~kHz}} 
\def\MHz{\mbox{~MHz}} 
\def\GHz{\mbox{~GHz}} 
\def\THz{\mbox{~THz}} 
\def\ns{\mbox{~ns}} 
\def\mus{~\mu\hbox{s}} 
\def\ms{\mbox{~ms}} 
\def\sec{\mbox{~s}} 
\def\min{\mbox{~m}}
\def\hr{\mbox{~h}}
\def\yr{\mbox{~yr}}
\def\Jy{\mbox{~Jy}}
\def\ms{{\mbox{~ms}}}
\def\astar{A$_{\star}$}
\def\Msun{M$_{\odot}$}
\def\swift{{\textit{Swift}}\xspace}
\def\fermi{{\textit{Fermi}}\xspace}

\def\gbm{{\textit{Fermi}-GBM}\xspace}
\def\lat{{\textit{Fermi}-LAT}\xspace}
\def\Epk{E$_{\textrm{peak}}$}
\def\Eiso{E$_{\textrm{iso}}$}
\def\Liso{L$_{\textrm{iso}}$}
\def\t90{T$_{\textrm{90}}$}
\def\tvar{$t_{\textrm{var}}$}
\def\t0{$t_{0}$}
\def\nufnu{$\nu F_{\nu}$}
\def\ra#1#2#3{#1$^{^\textrm{h}}$#2$^{^\textrm{m}}$#3$^{^\textrm{s}}$}
\def\dec#1#2#3{#1$^\circ$#2$'$#3$''$}
\def\grb{GRB 230307A }
\def\fluence{\textrm{erg}\cdot\textrm{cm}^{-2}}


\def\aa{A\&A} 
\def\aar{A\&A~Rev.} 
\def\aas{A\&AS} 
\def\actaa{Acta Astron.} 
\def\aj{AJ} 
\def\ao{Appl.~Opt.} 
\def\apj{ApJ} 
\def\apjl{ApJ} 
\def\apjs{ApJS} 
\def\apjsupp{ApJ Supp. Series} 
\def\aplett{Astrophys.~Lett.} 
\def\apspr{Astrophys.~Space~Phys.~Res.} 
\def\apss{Ap\&SS} 
\def\araa{ARA\&A} 
\def\arxiv{arXiv} 
\def\arxive{arXiv e-prints} 
\def\azh{AZh} 
\def\baas{BAAS} 
\def\bac{Bull. astr. Inst. Czechosl.} 
\def\bain{Bull.~Astron.~Inst.~Netherlands} 
\def\caa{Chinese Astron. Astrophys.} 
\def\cjaa{Chinese J. Astron. Astrophys.} 
\def\cqg{CQGrav} 
\def\ea{Exp. Astron.} 
\def\fcp{Fund.~Cosmic~Phys.} 
\def\gca{Geochim.~Cosmochim.~Acta} 
\def\gcn{GCN Circ. } 
\def\grl{Geophys.~Res.~Lett.} 
\def\iaucirc{IAU~Circ.} 
\def\icarus{Icarus} 
\def\jcap{J. Cosmology Astropart. Phys.} 
\def\jcp{J.~Chem.~Phys.} 
\def\jgr{J.~Geophys.~Res.} 
\def\jqsrt{J.~Quant.~Spec.~Radiat.~Transf.} 
\def\jrasc{JRASC} 
\def\memras{MmRAS} 
\def\memsai{Mem.~Soc.~Astron.~Italiana} 
\def\mnras{MNRAS} 
\def\na{New A} 
\def\nar{New A Rev.} 
\def\nat{Nature} 
\def\nphysa{Nucl.~Phys.~A} 
\def\nucinstrummethodsa{Nucl.~Instrum.~Methods~A} 
\def\pasa{PASA} 
\def\pasj{PASJ} 
\def\pasp{PASP} 
\def\physrep{Phys.~Rep.} 
\def\physscr{Phys.~Scr} 
\def\planss{Planet.~Space~Sci.} 
\def\pra{Phys.~Rev.~A} 
\def\prb{Phys.~Rev.~B} 
\def\prc{Phys.~Rev.~C} 
\def\prd{Phys.~Rev.~D} 
\def\pre{Phys.~Rev.~E} 
\def\prx{Phys.~Rev.~X} 
\def\prl{Phys.~Rev.~Lett.} 
\def\procspie{Proc.~SPIE} 
\def\psci{PoS} 
\def\qjras{QJRAS} 
\def\rmxaa{Rev. Mexicana Astron. Astrofis.} 
\def\skytel{S\&T} 
\def\solphys{Sol.~Phys.} 
\def\ssr{Space~Sci.~Rev.} 
\def\sovast{Soviet~Ast.} 
\def\zap{ZAp} 


\newcommand\aastex{AAS\TeX}
\newcommand\latex{La\TeX}


\definecolor{burntorange}{rgb}{0.8, 0.33, 0.0}
\definecolor{amber}{rgb}{1.0, 0.75, 0.0}
\definecolor{ao(english)}{rgb}{0.0, 0.5, 0.0}
\definecolor{darkorchid}{rgb}{0.6, 0.2, 0.8}
\definecolor{aqua}{rgb}{0.0, 1.0, 1.0}
\definecolor{brightlavender}{rgb}{0.75, 0.58, 0.89}

\newcommand{\red}[1]{\textcolor{red}{#1}}
\newcommand{\orange}[1]{\textcolor{burntorange}{#1}}
\newcommand{\yellow}[1]{\textcolor{amber}{#1}}
\newcommand{\green}[1]{\textcolor{ao(english)}{#1}}
\newcommand{\blue}[1]{\textcolor{blue}{#1}}
\newcommand{\purple}[1]{\textcolor{darkorchid}{#1}}

\newcommand{\hlr}[1]{\sethlcolor{pink}{\hl{#1}}\sethlcolor{yellow}}
\newcommand{\hlo}[1]{\sethlcolor{orange}{\hl{#1}}\sethlcolor{yellow}}
\newcommand{\hly}[1]{\sethlcolor{yellow}{\hl{#1}}\sethlcolor{yellow}}
\newcommand{\hlg}[1]{\sethlcolor{green}{\hl{#1}}\sethlcolor{yellow}}
\newcommand{\hlb}[1]{\sethlcolor{aqua}{\hl{#1}}\sethlcolor{yellow}}
\newcommand{\hlp}[1]{\sethlcolor{brightlavender}{\hl{#1}}\sethlcolor{yellow}}

\newcommand{\add}[1]{\textcolor{blue}{#1}}
\newcommand{\cut}[1]{\textcolor{blue}{\st{#1}}}


\NewDocumentCommand\bullets{>{\SplitList{;}}m}
  {
    \begin{itemize}
      \ProcessList{#1}{ \insertitem }
    \end{itemize}
  }
\NewDocumentCommand\numbers{>{\SplitList{;}}m}
  {
    \begin{enumerate}
      \ProcessList{#1}{ \insertitem }
    \end{enumerate}
  }
\newcommand\insertitem[1]{\item #1}

\newcommand{\addref}{\red{(add reference)}\xspace}
\newcommand{\addrefs}{\red{(add references)}\xspace}
\newcommand{\addequ}{\red{(add equation)}\xspace}
\newcommand{\addim}{\red{(add image)}\xspace}


\newcommand{\tte}{\texttt{TTE}\xspace}
\newcommand{\ctime}{\texttt{CTIME}\xspace}
\newcommand{\cspec}{\texttt{CSPEC}\xspace}
\newcommand{\drm}{\texttt{DRM}\xspace}
\newcommand{\drms}{\texttt{DRM}s\xspace}
\newcommand{\lle}{\texttt{LLE}\xspace}

\newcommand{\DistortedCountsSpectrum}{\texttt{dCntsSpec}\xspace}
\newcommand{\AssumedModelSpectrum}{\texttt{ModelSpec}\xspace}
\newcommand{\AssumedModelCountsSpectrum}{\texttt{ModelCntsSpec}\xspace}
\newcommand{\DistortedAssumedModelCountsSpectrum}{\texttt{dModelCntsSpec}\xspace}

\def\targeted{\texttt{targeted search}\xspace}
\def\untargeted{\texttt{untargeted search}\xspace}

\def\pom{$\pm$\xspace}
\def\abt{$\sim$}

\newcommand{\kTmin}{$kT_{\textrm{min}}$\xspace}
\newcommand{\kTmax}{$kT_{\textrm{max}}$\xspace}

\newcommand{\Cstat}{$C_{\textrm{stat}}$\xspace}
\newcommand{\PGstat}{$PG_{\textrm{stat}}$\xspace}
\newcommand{\Pstat}{$P_{\textrm{stat}}$\xspace}
\newcommand{\chisq}{$\chi^{2}$\xspace}
\newcommand{\chisqdof}{$\chi^{2}_{\nu}$\xspace}

\definecolor{lightgray}{gray}{0.9}
\definecolor{darkgray}{gray}{0.7}

\def\checkmark{\tikz\fill[scale=0.4](0,.35) -- (.25,0) -- (1,.7) -- (.25,.15) -- cycle;}

%
\newcommand{\CNRS}{\affiliation{Universit\'e Paris-Saclay, CNRS/IN2P3, IJCLab, 91405 Orsay, France}}
\newcommand{\CSPAR}{\affiliation{Center for Space Plasma and Aeronomic Research, University of Alabama in Huntsville, Huntsville, AL 35899, USA}}
\newcommand{\FIT}{\affiliation{Department of Aerospace, Physics and Space Sciences, Florida Institute of Technology, Melbourne, FL 32901, USA}}
\newcommand{\GSFC}{\affiliation{NASA Goddard Space Flight Center, University of Maryland, Baltimore County, Greenbelt, MD 20771, USA}}
\newcommand{\GSFCAstro}{\affiliation{Astrophysics Science Division, NASA Goddard Space Flight Center, Greenbelt, MD 20771, USA}}
\newcommand{\Amentum}{\affiliation{Amentum Space Exploration Group, Huntsville, AL 35806, USA}}

\newcommand{\INAFOAR}{\affiliation{INAF-Osservatorio Astronomico di Roma, Via Frascati 33, 00076 Monte Porzio Catone (RM), Italy}}

\newcommand{\LSU}{\affiliation{Department of Physics and Astronomy, Louisiana State University, Baton Rouge, LA 70803 USA}}
\newcommand{\LosAlamos}{\affiliation{Center for Non Linear Studies, Los Alamos National Laboratory, Los Alamos, NM, 87545, USA}}
\newcommand{\MPI}{\affiliation{Max-Planck-Institut f\"{u}r extraterrestrische Physik, Giessenbachstrasse 1, D-85748 Garching, Germany}}
\newcommand{\MSFCAstro}{\affiliation{ST12 Astrophysics Branch, NASA Marshall Space Flight Center, Huntsville, AL 35812, USA}}
\newcommand{\NASA}{\altaffiliation{NASA Postdoctoral Fellow}}
\newcommand{\NYUAbuDhabi}{\affiliation{Center for Astro, Particle, and Planetary Physics, New York University Abu Dhabi}}
\newcommand{\PdiB}{\affiliation{Dipartimento Interateneo di Fisica dell'Università e Politecnico di Bari, Via E. Orabona 4, 70125, Bari, Italy}}
\newcommand{\SdiB}{\affiliation{Istituto Nazionale di Fisica Nucleare - Sezione di Bari, Via E. Orabona 4, 70125, Bari, Italy}}
\newcommand{\SPA}{\affiliation{Department of Space Science, University of Alabama in Huntsville, 320 Sparkman Drive, Huntsville, AL 35899, USA}}
\newcommand{\SRON}{\affiliation{SRON Netherlands Institute for Space Research, Niels Bohrweg 4, 2333CA Leiden, The Netherlands}}
\newcommand{\UAH}{\affiliation{University of Alabama in Huntsville, 320 Sparkman Drive, Huntsville, AL 35899, USA}}
\newcommand{\UCD}{\affiliation{School of Physics, University College Dublin, Belfield, Dublin 4, Ireland}}
\newcommand{\USRA}{\affiliation{Science and Technology Institute, Universities Space Research Association, Huntsville, AL 35805, USA}}
%

%
\author[0000-0003-1835-570X]{S.~Dalessi}
\SPA
\CSPAR
\email{sd0104@uah.edu}
\author[0000-0002-2149-9846]{P.~Veres}
\SPA
\CSPAR
\email{peter.veres@uah.edu}
\author[0000-0002-0468-6025]{C.~M.~Hui}
\MSFCAstro
\email{c.m.hui@nasa.gov}

\author[0000-0002-6657-9022]{S.~Bala}
\USRA
\email{sumanbala2210@gmail.com}

\author[0000-0001-8058-9684]{S.~Lesage}
\SPA
\CSPAR
\email{sjl0014@uah.edu}
\author[0000-0003-2105-7711]{M.~S.~Briggs}
\SPA
\CSPAR
\email{briggsms@uah.edu}
\author[0000-0002-0587-7042]{A.~Goldstein}
\USRA
\email{AGoldstein@usra.edu}
%
%
\author[0000-0002-2942-3379]{E.~Burns}
\LSU
\email{ericburns@lsu.edu}
\author[0000-0002-8585-0084]{C.~A.~Wilson-Hodge}
\MSFCAstro
\email{colleen.wilson@nasa.gov}
\author[0000-0002-0186-3313]{C.~Fletcher}
\USRA
\email{cfletcher@usra.edu}

\author[0000-0002-7150-9061]{O.~J.~Roberts}
\USRA
\email{oliver.roberts@nasa.gov}

\author[0000-0001-7916-2923]{P. N. ~Bhat}
\CSPAR
\email{bhatn@uah.edu}


\author[0000-0001-9935-8106]{E.~Bissaldi}
\PdiB
\SdiB
\email{Elisabetta.Bissaldi@ba.infn.it}

\author[0009-0003-3480-8251]{W.~H.~Cleveland}
\USRA
\email{william.cleveland@nasa.gov}
\author{M.~M.~Giles}
\Amentum
\email{misty.m.giles@nasa.gov}

\author[0000-0002-2926-0469]{M.~Godwin}
\SPA
\CSPAR
\email{msg0028@uah.edu}
\author[0000-0003-0761-6388]{R.~Hamburg}
\USRA
\email{rhamburg@usra.edu
}
\author[0000-0001-9556-7576]{B.~A.~Hristov}
\CSPAR
\email{bah0046@uah.edu}

\author[0000-0001-9201-4706]{D.~Kocevski}
\MSFCAstro
\email{daniel.kocevski@nasa.gov}

\author[0000-0002-2531-3703]{B.~Mailyan}
\FIT
\email{mbagrat@gmail.com}

\author[0000-0002-0380-0041]
{C.~Malacaria}
\INAFOAR
\email{cmalacaria.astro@gmail.com}

\author[0000-0002-4593-0792]{O.~ Mukherjee}
\USRA
\email{omukherjee@usra.edu}

\author[0000-0002-0602-0235]{L.~Scotton}
\SPA
\CSPAR
\email{ls0162@uah.edu}

\author[0000-0002-0221-5916]{A.~von Kienlin}
\MPI
\email{azk@mpe.mpg.de}
\author[0000-0001-9012-2463]{J.~Wood}
\MSFCAstro
\email{joshua.r.wood@nasa.gov}



\begin{abstract}
On March 7th, 2023 the \textit{Fermi} Gamma-ray Burst Monitor observed the second highest fluence gamma-ray burst (GRB) ever, GRB~230307A. With a duration beyond 100~s, GRB~230307A contains a multitude of rapidly-varying peaks, and was so bright it caused instrumental effects in the GBM detectors. The high fluence of this burst, (6.02 $\pm$ 0.02)$\times$10$^{-3}$ erg cm$^{-2}$,  prompted rapid follow-up across the electro magnetic spectrum including the discovery of an associated kilonova. GRB~230307A is one of a few long GRBs with an associated compact merger origin. Three main temporal regions of interest are identified for fine time-resolution spectral analysis: triggering pulse, main emission, and late emission, and the parameter evolution is traced across these regions. The high flux of the burst allowed for the statistical preference of a more complex, physically-motivated model, the Double Smoothly Broken Power Law, over typical spectral fitting functions for GRBs. From this model the evolution of the parameters was found to be in accordance with those expected for synchrotron radiation in the fast-cooling regime. Additionally, it was found that the flux experiences a steep decline in late time intervals, a feature which is often attributed to high-latitude emission, which follows the dissipation episodes. Furthermore, GRB~230307A was found to have one of the highest inferred bulk Lorentz factors of $\Gamma = 1600$. GRB~230307A is a noteworthy burst in terms of flux alone, but additionally provides a unique insight into the possible temporal and spectral characteristics of a new long merger class of GRBs.

\end{abstract}


\section{Introduction}
Gamma-Ray Bursts (GRBs) \citep[first reported in ][]{discovery} are the most luminous explosions in the Universe and are usually divided into two groups based on their duration of prompt emission
phase \citep{Dezalay1992, kouveliotou93}. Short-duration (<2~s) GRBs are predominantly produced by the merger of two compact objects, such as binary neutron stars (BNS) mergers \citep{narayan92, Thompson94, Goldstein, Abbott_170817A, Fong+15sGRB} or neutron star-black hole (NS-BH) mergers \citep{ eichler89, nakar07}, and long-duration (>2~s) GRBs originate from a subtype of core collapse of massive stars \citep{woosley93, paczynski98, macfadyen99, woosley06}.
The duration distributions overlap, so occasionally the spectral information or the hardness ratio (the ratio of high-energy to low-energy flux) is used to distinguish between the type of GRBs \citep{Paciesas+99cat, Bhat_2016, Kienlin2020}. \cite{kouveliotou93} showed that the hardness ratio of the GRBs is anti-correlated with their duration; i.e short GRBs (sGRBs) are relatively harder and long GRBs (lGRBs) are observationally softer. 
These two parameters are not sufficient to conclude the physical origin of GRBs alone, but serve as possible signifiers see e.g. \citep{Zhang+09typeI, Kann2011}. 
On the other hand, lGRBs are reliably associated with Supernovae Ic-BL \citep{Hjorth2003} and sGRBS are associated with kilonovae \citep{Tanvir2013, Yang_2015, Abbott+17BNS, Levan_2023, Yang2024}.
Another interesting feature of GRBs is the variability time-scale. Constraints on the size of the emitting region can be placed based on causality arguments \citep{rybicki79}, using the variability time-scale.
The typical short variability time-scale can be around 10~ms, but in a few cases variability < 10~ms has been observed \citep{Veres_2023}. In the most extreme case, a $\sim200$~\textmu s variation \citep{Bhat+92shortvar} has been reported. Generally, sGRBs have shorter variability than lGRBs \citep{Bhat+12mvt, Golkhou+15tvar}. 

The temporal difference between a GRB's light curves in different energy bands can also help to categorize between GRBs \citep{gehrels06, Zhang+06ag, norris06}. The spectral lag is defined as positive when high-energy photons precede low-energy photons. In general, lGRBs show a positive spectral lag \citep{1995_Cheng, 1997_band, Norris+00lag, Ukwatta+10lag}, while sGRBs are characterized by zero spectral lag \citep{Norris+00lag, Ukwatta+10lag}. In very few cases, GRBs show negative spectral lag, such as GRB 090426C and GRB 150213A, \citep{CHAKRABARTI_2018}, indicating a soft-to-hard transition. An anti-correlation has also been observed between the spectral lag and the peak luminosity (lag-luminosity relation) \citep{norris00, norris02, Ukwatta+10lag}. This correlation indicates that spectral lags could be used to determine physical luminosities of GRBs.

A few GRBs have made the picture more complicated. Previously  GRB~060614 stood alone as an outlier, as it had long duration of $\sim$102~s but the peak luminosity and temporal lag fell within the short-duration GRB subclass \citep{gehrels06}. Also, no supernova counterpart was found down to very deep optical limits.  \cite{Ahumada_2021} discovered the shortest lGRB (T$_{90}\sim$1.14~s), GRB 200826A, for which optical and X-ray follow-up observations confirmed an association with a collapsar.
Conversely, GRB 211211A with duration T$_{90} = 34.3 \pm 0.6$ ~s was found to have a kilonova counterpart with a luminosity, duration, and color similar to that which accompanied the  BNS merger detected in gravitational waves GW170817 \citep{Troja+22grb211211a,Rastinejad+22grb211211a}. GRB 211211A also had a short variability time-scale \citep{Veres_2023}. It is clear that measures of duration and hardness ratio are not sufficient to classify the progenitors of GRBs.

GRB 230307A was an exceptional burst, first reported by the \gbm (GBM) and was quickly identified to be one of the brightest GRBs ever observed \citep{GCN_33414}. Aside from its large flux, this GRB has many unique features.\cite{Dichiara_2023} have interpreted the initial soft pulse as a bright precursor. They also claimed the central engine to be a rapidly rotating magnetar with magnetic field $>10^{15}$G. An achromatic temporal break in the high-energy band during the prompt emission phase was reported by \cite{sun2023magnetar}. 
They claimed the presence of break reveals a narrow jet with a half opening angle of approximately 3.4$^\circ$. With the James Webb Space Telescope mid-infrared imaging and spectroscopy, \cite{Levan_2023} showed the presence of a kilonova similar to AT2017gfo, associated with GW170817. They also reported the likely identification of an atomic line signature of Tellurium and report a kilonova peak time in infrared at 30 days, both indicative of rapid neutron-capture nucleosynthesis. This result is also supported by \cite{gillanders_2023} and \cite{Yang2024} utilizing additional observations. These studies indicate that GRB 230307A belongs to the class of lGRBs associated with compact binary mergers. 

In this work, we present our analysis of the \gbm data of GRB 230307A and compare its properties with other GRBs observed by Fermi. In Section 2 we explained the details of the observation and our data selection procedure. In Section 3 we present our temporal and spectral analysis. Section 4 contains discussion and Section 5 our conclusion. \par

\label{sec:introduction}

\section{Observation} \label{sec:observation}

\grb (GBM burst number230307656) triggered the \gbm flight software on 2023 March 7 at 15:44:06.67 UTC ($t_0$). \gbm distributed an automated localization through the General Coordinates Network (GCN). The extraordinarily high flux of the burst was first noted in a GCN by GECAM \citep{GCN_33406}. A secondary manual GCN Circular was sent by the GBM team to notify the community of this event and encourage follow-up across all wavelengths \citep{GCN_33407}. Some notable results from follow-up efforts include: 
\begin{itemize}
    \item Multiple rounds of improved localization from the InterPlanetary Network leading to the successful follow-up observations \citep{GCN_33413, GCN_33425, GCN_33461}. 
    \item A redshift of 0.065 as first reported by \cite{GCN_33485}. 
    \item Independent observations from the Solar Orbiter STIX \citep{GCN_33410}.
    \item Upper limits of neutrino flux from IceCube \citep{GCN_33430}.
    \item Detection of late time X-ray afterglow by Chandra \citep{GCN_33558}.
    \item Serendipitous coverage by TESS and LEIA providing prompt optical and X-ray coverage of a merger for the first time \citep{GCN_33453, GCN_33466}.
    \item  Two rounds of observations by the James Webb Space Telescope, confirming an associated kilonova and favoring the nearby distance of the event \citep{GCN_33569, GCN_33747}.

\end{itemize}

\gbm is one of two science instruments onboard the \textit{Fermi Gamma-ray Space Telescope}, the other being the \textit{Fermi} Large Area telescope (LAT). \gbm was designed to detect and localize bursts in the 8 keV to 40 MeV range \citep{Meegan_2009}. \gbm consists of an array of twelve Sodium Iodide  (NaI) and two Bismuth Germanate (BGO) detectors for prompt detection of GRBs. The NaI detectors are placed at different orientations around the spacecraft to observe the entire unocculted sky in the 8~keV to 1000~keV energy range. The two BGO detectors observe the 200~keV to 40~MeV energy range and are placed on opposite sides of the spacecraft. 

\gbm produces three sets of data products at differing resolutions that can be used for data analysis: Time-tagged Event (\texttt{TTE}), Continuous Spectroscopy (\texttt{CSPEC}), and Continuous Time (\texttt{CTIME}) data. Both the \texttt{TTE} and \texttt{CSPEC} data sets have a full 128 spectral channel resolution, the \texttt{TTE} has the highest resolution at 2 microseconds, while \texttt{CSPEC} is binned at 1.024~s.

\subsection{Data Handling}

The high photon flux produced by GRB~230307A created time periods with data issues (i.e. bad time intervals; BTIs)\footnote{\url{https://fermi.gsfc.nasa.gov/ssc/data/analysis/grb230307a.html}\label{ss:btis}}, in \gbm data \citep{GCN_33551}. Binned (\texttt{CSPEC} and \texttt{CTIME}) and unbinned (\texttt{TTE}) \gbm data types are affected slightly differently due to how the on-board electronics processes these data types.

Unbinned data experiences issues when the summed count rate of all detectors exceeds the 375\,kHz data rate limit of the \gbm high-speed science data bus. Beyond this limit, \texttt{TTE} telemetry packets are lost and the data are irrecoverable \citep{Meegan_2009}. The unbinned data loss  results in an incorrect inference on brightness, but does not affect spectral behavior.  Although this effect is not present for the full BTI of this GRB, it did occur in a few brief instances between \t0+3 and \t0+7 seconds. Binned data does not experience this same irrecoverable packet loss.

For all \gbm data types, higher than normal count rates create dead time which is automatically corrected by the software before generating the resulting \gbm FITS files. This technique is only valid when a single detector experiences input count rates below $\sim$60k counts per second (cps). Above this threshold, more complex dead time and pulse pile-up (PPU) effects occur \citep{Meegan_2009}. As explained in \cite{Chaplin2013} and \cite{Bhat2014}, \gbm data beyond the $\sim$60k~cps PPU regime distorts both the observed spectral shape and intensity of the data. As reported in \cite{GCN_33551}, GRB~230307A experiences PPU between \t0+2.752~s to \t0+10.944~s. Even with mild PPU, as is the case for GRB~230307A, analyzing the data without correction will misrepresent the true spectrum of the event. A more detailed description of these effects on \gbm data and how to properly correct them can be found in \cite{Lesage_2023}.

\subsection{GBM Data}
\grb was observed by \gbm with significant signal from \t0 until $t_0 + 95.770$~s and was visible until  $t_0 + 128$~s when it was occulted by the Earth.  The period of time during which 90\% of the emission took place, T$_{90}$, is $34.56\pm 0.57$~s. Despite signal being present in all 12 NaI detectors and both BGO detectors, due to the orientation of the burst coming through the bottom of the spacecraft, only detector {\tt NA} has a detector-source angle of less than $60 \deg$. Detector {\tt NB} has a slightly larger detector-source angle ($61 \deg$), but is actually blocked by detector {\tt NA}. Any other detectors had too large detector-source angles or were blocked by the spacecraft itself, so only {\tt NA} and {\tt B1} are used for spectral analysis.  For the duration of the detection until occultation, the incident angle with respect to  \textit{Fermi-}LAT was $142\deg$.

Preliminary spectral analysis and calculation for the T$_{90}$ were done using the {\tt RMfit}\footnote{\url{https://fermi.gsfc.nasa.gov/ssc/data/analysis/rmfit}} software and reported in a GCN Circular \citep{GCN_33411}. Fine time spectral analysis was conducted using the {\tt GBM Data Tools} \footnote{\url{https://fermi.gsfc.nasa.gov/ssc/data/analysis/gbm/}} software. Energy channels in the range of 8-900~keV for the NaI detectors and 0.3-39~MeV for the BGO detectors were selected. Additionally, the energy channels between 30-40~keV were excluded due to the Iodine K-edge at 33.17~keV that can cause significant residuals for bright sources \citep{Meegan2009}. 

\begin{figure*}[!hbt]
    \centering
    \includegraphics[width=1\textwidth, height=0.8\textwidth]{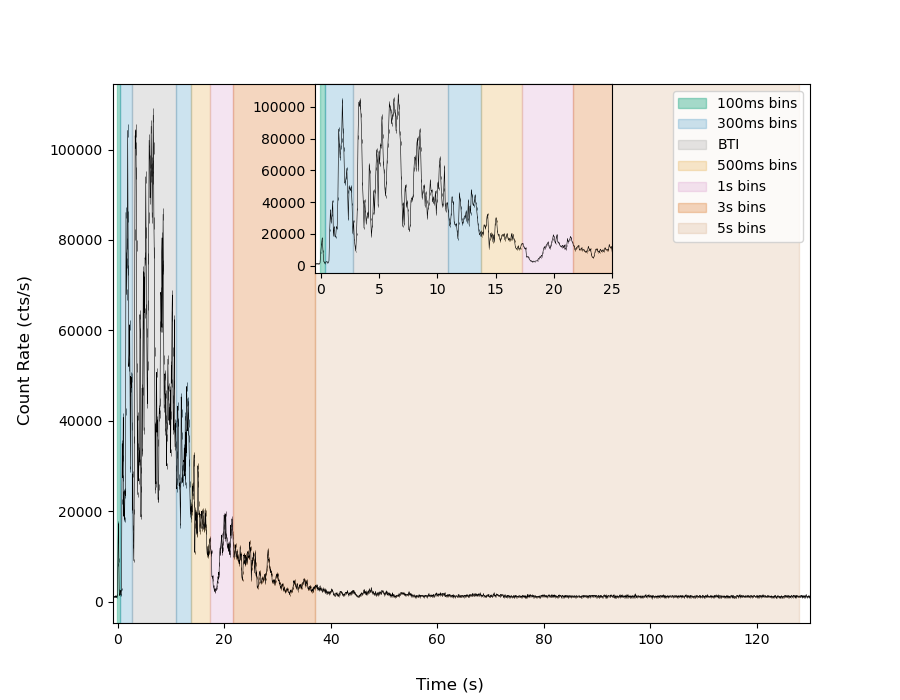}
    \caption{Plot of the \grb lightcurve of the NaI detector with variable signal-to-noise (SNR) binning minimum widths. The different colored regions represent the different temporal resolutions of the bins.}
    \label{fig:Fig 1}
\end{figure*}

\section{Time Resolved Analysis} \label{sec:time_resolved}
 
The lightcurve for \grb can be seen in Figure 1, where the grey-shaded regions represent the BTI. As opposed to some GRBs with easily identifiable simple pulse structures, \grb exhibits a very complex lightcurve with many pulses. \cite{Hakkila_Preece_14} note that GRBs with a multitude of pulses often signify rapidly-varying emission. Instead of a monotonic overall hard-to-soft evolution, these many peaks are seen to be the results of embedded relativistic shock structures.

Due to this burst's extraordinarily high count rate, it is possible to conduct fine-time spectral analysis to track the evolution of the spectral parameters throughout the burst. To determine the time intervals, the TTE data was rebinned based on the signal-to-noise ratio (SNR) of 100 for the NaI detectors. The SNR=100 was chosen, as to best match the pulses based on visual inspection. The temporal binning was based on using the {\tt NA} lightcurve in the standard GRB 50-300~keV energy range, as it encompasses the total spectral evolution across the full duration of the burst. A minimum bin width constraint of 100~ms was placed to ensure a sufficient number of counts per bin to constrain spectral fits in order to capture spectral evolution. If spectral parameters could not be constrained within a 30\% error, the minimum bin width was increased by increments of 200~ms and then 2~s for later time bins. Spectral modeling was performed with a minimum temporal bin constraint to ensure sufficient photon counts, as SNR can be insufficient for time-resolved spectroscopy due to background fit fluctuations. The definition of naming convention for periods as well the time ranges that correspond to which minimum bin width can be found in Table \ref{tab:snr_tabel} and is visually represented in Figure \ref{fig:Fig 1}.

\begin{table}[]
    \centering
    \footnotesize
    \begin{tabular}{c|c|c}
    Interval    & Minimum bin duration & Time range (s)  \\ \hline
    Triggering Pulse  & 100~ms & -0.064-0.355 \\  \hline
    Main Emission     & 300~ms  & 0.355-2.752 \\
                BTI      & CSPEC- 1.024~s & 2.752-10.944 \\
                      & 300~ms  & 10.944-13.712 \\
                      & 500~ms  & 13.712-17.264 \\  \hline 
    Dip & 500~ms & 17.264 - 19.604   \\ \hline               
    Late Emission                       & 1~s  & 19.604- 21.652 \\        
                        & 3s  & 21.652- 36.965 \\ \hline
    Tail              & 5~s & 36.965-128.000 \\ \hline
    \end{tabular}
    \caption{Definition of periods of interest and the minimum bin duration for specified time ranges, relative to \t0. }
    \label{tab:snr_tabel}
\end{table}

\subsection{Spectral Fitting}

While conducting preliminary spectral analysis it was found that the standard one-component models used in the GBM Spectral catalogs \citep{Poolakkil_GBM_10yr}: Band, Compton, Power Law, and Smoothly Broken Power Law, were insufficient in describing the full spectrum of this burst. These models either returned  unconstrained fit parameters or residuals greater than $3\sigma$. Therefore, a selection was made to use the Double Smoothly Broken Power Law (2SBPL) that \citealt{Ravasio_18} defined as: 

\begin{equation}
\footnotesize
\begin{split}
N^{\rm 2SBPL}_{\rm E} = A \, E_{\rm break}^{\alpha_1} \, \Biggl[ \, \Biggl[ \Biggl(\frac{E}{E_{\rm break}}\Biggr)^{- \alpha_1 n_1}+\Biggl(\frac{E}{E_{\rm break}}\Biggr)^{- \alpha_2 n_1}\Biggr]^{\frac{n_2}{n_1}}\\
+\Biggl(\frac{E}{E_{\rm j}}\Biggr)^{- \beta \, n_2} \cdot \Biggl[\Biggl(\frac{E_{\rm j}}{E_{\rm break}}\Biggr)^{- \alpha_{1} n_{1}}+\Biggl(\frac{E_{\rm j}}{E_{\rm break}} \Biggr)^{- \alpha_2 n_1} \Biggr]^{\frac{n_2}{n_1}} \Biggr]^{-\frac{1}{n_2}}~,
\end{split}
\label{eq:2sbpl}
\end{equation}
where
\begin{equation}
\footnotesize
E_{\rm j} = E_{\rm peak} \cdot \Biggl(- \frac{\alpha_2 + 2}{\beta + 2}\Biggr) ^{\frac{1}{(\beta - \alpha_2) \, n_2}}~.
\end{equation}
\\
The free parameters for this form are the amplitude $A$, the break energy $E_{break}$, the peak energy $E_{peak}$, the photon index below the break $\alpha_1$, the photon index between the break and the peak $\alpha_2$, the high energy photon index $\beta$, and the smoothness parameters $n_1$ and $n_2$. The smoothness parameter $n_1= 5.38$ was fixed for the break energy, which corresponds to a sharper curvature around the break and was the mean value of the distribution when $n_1$ was left to vary in \citealt{Ravasio_18}. Additionally, $n_2=2.69$ was fixed for the curvature around the peak energy which is derived from the smoothness parameter $\Lambda =0.3$ used in the \gbm catalog \citep{Kaneko+06batse}. 
Furthermore, the constraint was added that $\alpha_1 > \alpha_2$ to aid in the convergence of the model fit parameters and be in accordance with the physical interpretation of the 2SBPL, and $E_{break}$ > 10~keV to match the minimum bandpass limit of \gbm. 

\subsection{Triggering Pulse}
The initial triggering pulse (\t0-0.064 - \t0+0.355~s), is more prominent in the lower energies (8-300~keV) as shown in Figure 2. 
A further study of this triggering pulse by \citealt{Dichiara_2023}, found that the lower flux, soft spectrum, and delay with respect to the onset of the main emission were consistent with features of a GRB precursor.  They found that the spectra for this precursor was best fit by a Band model, although a more complex Band + Blackbody fit was also well constrained, but was not statistically preferred to the Band model fit. Spectral fitting conducted in our study over this time range using Band, Band+Blackbody, and Multicolor Blackbody did not show sufficient evidence for a thermal component. 

\subsection{Main Emission and Pulse Pile-Up Correction }
The main emission of the burst lasts from \t0+0.355-\t0+17.264~s and covers the brightest part of the burst consisting of multiple pulses. Due to the high variability and multitude of peaks, the \texttt{TTE} data is binned to 300~ms for the beginning of the BTIs and switched to 500~ms as the overall count rate and variability begins to decrease (Table \ref{tab:params}). In the PPU region, a correction was applied to \texttt{CSPEC} data assuming the 2SBPL model. These corrections resulted in an increase of the photon flux on the order of 5\%. Spectral fits in the BTI region should be treated with caution as they were made under the assumption that the 2SBPL is the true spectrum. The PPU correction is model-dependent and may result in a different flux if a photon model other than 2SBPL is used.

\subsection{The Dip}
\begin{figure}
    \includegraphics[width=1.1\columnwidth,height=1.15\columnwidth]{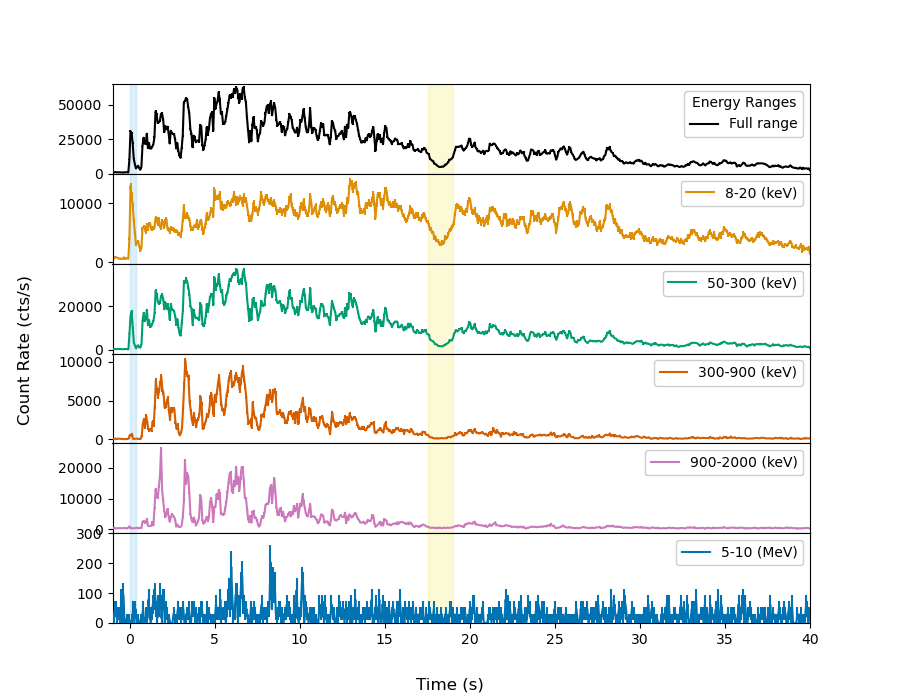}
    \caption{Lightcurve of \grb split in five different energy ranges. The top panel shows the full energy range, while the second and third panels cover the energy ranges observed by the NaI detectors. The fourth, fifth, and sixth panels show the energy ranges of the BGO detectors. The two highlighted regions indicate the triggering pulse and dip respectively. }
    \label{fig:lc_energy}
\end{figure}

Beyond the primary emission episode, an additional period of interest is identified for further analysis: the parabolic  ``dip" (\t0+17.264~s - \t0+ 19.604~s) (Figure \ref{fig:lc_energy}). The dip is a feature that is not only present in the \gbm lightcurve, but is seen in the observations of AGILE/MCAL \citep{GCN_33412,GCN_33444}, GRBAlpha \citep{GCN_33418}, VZLUSAT-2 \citep{GCN_33424}, Konus-Wind \citep{GCN_33427}, AstroSat \citep{GCN_33437} and NuSTAR \citep{GCN_33478}. Therefore, the dip is not likely due to instrumental effects and is instead a property of \grb itself.

The dip exhibits persistent temporal features across a wide range of energies. In the \gbm data, the dip is clearly prominent as low as 8~keV and up to around 2~MeV as highlighted in  Figure \ref{fig:lc_energy}. However, the dip is not present in the LEIA lightcurve in the 0.5-4~keV range \citep{sun2023magnetar}. Due to the highly symmetrical appearance of the dip, a simple parabola is a good fit for the data across all energy ranges and observations. Using the bounds set by the time bins established by the SNR, the dip was found to have a duration of 2.62~s. In the 8-50~keV range the parabola has a curvature coefficient (represent the percentage change in the count rates) of 58.5 $\pm$ 7.1 \%, in the 50-300~keV range the curvature is 78.53$\pm$ 5.2 \%, in the 300-900~keV range the curvature is 20.5 $\pm$ 1.3 \%, and in the 900-2000~keV range the curvature is 15.1 $\pm$ 4.1\%. 
The largest change is seen in the 50-300~keV range with and almost 80\% decrease in counts during the dip. 

\subsection{Late Emission and Tail}
The late emission of the burst lasts from \t0+19.02-\t0+95.770~s with a significant signal. There is a weaker tail extending out to 128~s when the burst was occulted by Earth. During the late emission, there are a few smaller peaks and variability until around 30 seconds, and then there is a steady decay. Due to the high flux of this burst and subsequent scaling of the lightcurves this time region in Figure \ref{fig:Fig 1} appears relatively flat, but it is still well above the count rates seen for a typical GRB.

\section{Discussion} \label{sec:discussion}

\begin{figure*}[hp!]
    \centering
    \includegraphics[width=1.05\textwidth, height=1.25\textwidth]{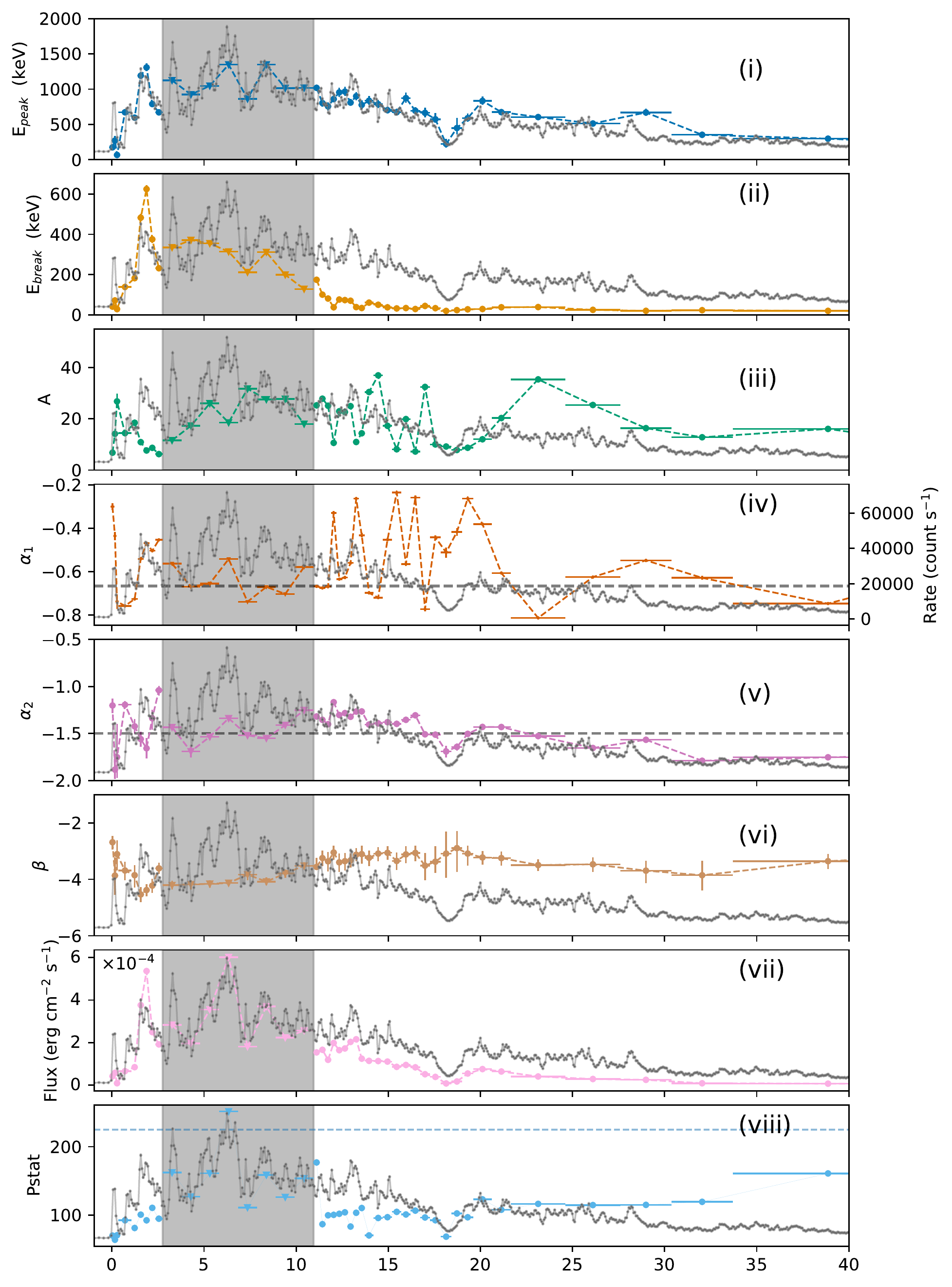}
    \caption{Spectral parameters of the 2SBPL for each of the 45 source intervals compared to the lightcurve. Horizontal bars represent the duration of the time range for which the spectral fit was conducted. Each of the parameters is presented in their own panel: (i) is the peak energy; (ii) is the break energy  
    (iii) is the normalization constant; (iv-vi) are the three photon indices where the horizontal dashed lines indicate the expected values for synchrotron emission; (vii) is the calculated flux; and (vii) is the fitting statistic from the likelihood function where the horizontal dashed line represents the degrees of freedom.}
    \label{fig:full_func}
\end{figure*}

\subsection{Evolution of Spectral Parameters}
\begin{figure*}[ht]

    \centering
    \includegraphics[width=0.8\textwidth, height=0.4\textwidth]{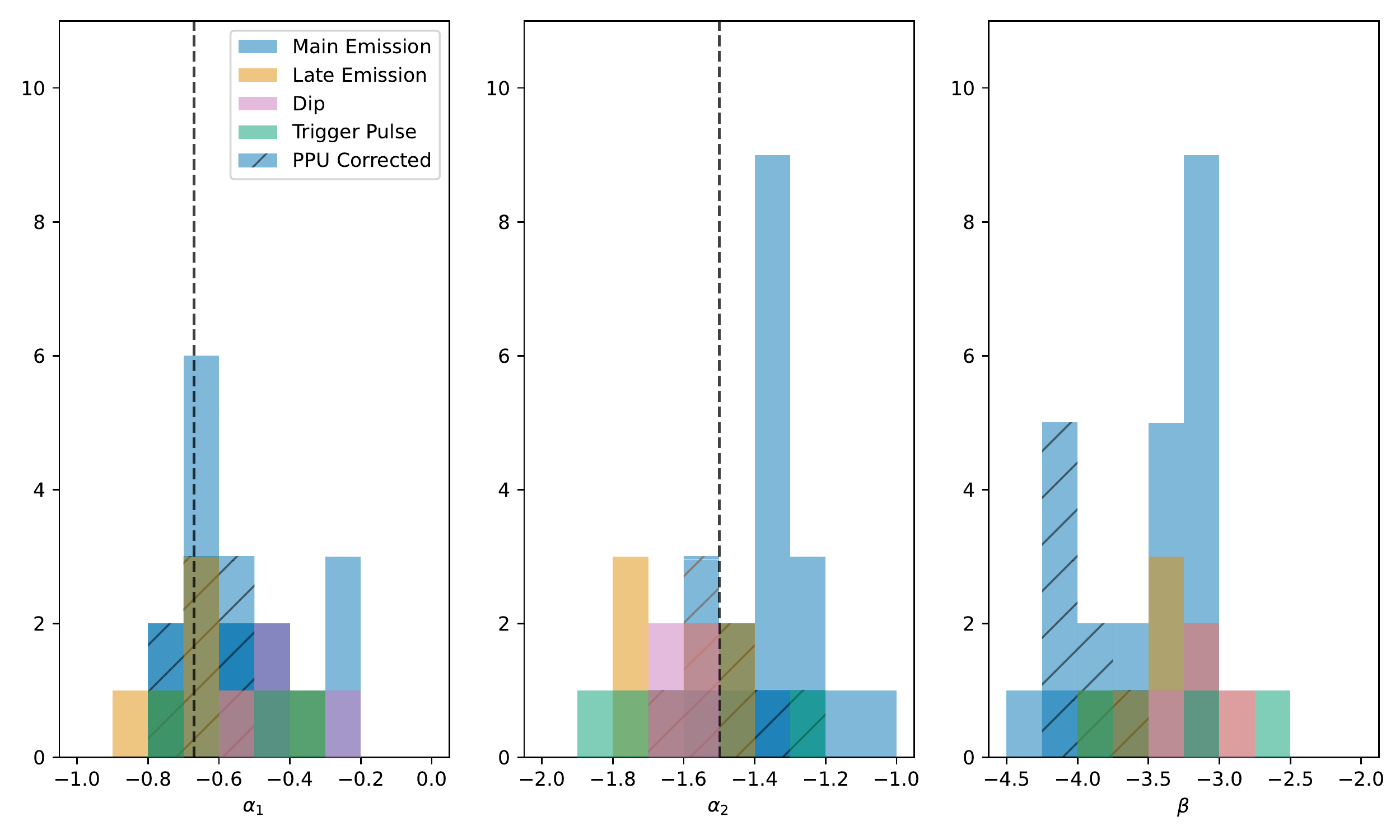}
    \caption{Distribution of $\alpha_1$, $\alpha_2$ and $\beta$ parameters for all spectral fits. Vertical dashed lines represent the predicted values for synchrotron emission of $\alpha_1 = -2/3$ and $\alpha_2 = -3/2$ in the fast cooling regime.}
    \label{fig:param_hist}
\end{figure*}

The results of the fine time spectral analysis are shown in Figure \ref{fig:full_func}, with the spectral parameters overplotted on the lightcurve of \grb to show the spectral evolution. The gray shaded region represents the BTI and the pulse pile-up corrected spectral values are also shown. The $E_{peak}$ parameter tracks the lightcurve, with a maximum of $E_{peak}= 1348^{+28}_{-25} $ keV during the 5.824-6.848s time bin. This maximum value is in agreement with the peak energy of $1321^{+60}_{-62} $ keV as reported by Konus-Wind \citep{GCN_33427}. 
 $E_{break}$ reaches a maximum of $E_{break}=624.2^{+20.8}_{-20.2}$ keV in the 1.726-2.034~s time bin and then decreases as a function of time.
 This evolution is consistent with the observed behavior of these parameters \citep[e.g.][]{Lu+12hard2soft}.  $E_{peak}$ tracks the photon count rate (intensity tracking), and $E_{break}$ shows the overall hard to soft spectral evolution. Notably, during the dip region, there is a decrease and then subsequent increase for both $E_{peak}$ and $E_{break}$.

Figure \ref{fig:param_hist} shows the distribution of the $\alpha_1$, $\alpha_2$, and $\beta$ parameter values for the different regions of the burst. The values for $\alpha_1$ are distributed around the expected value of -2/3 as predicted for synchrotron emission, with a mean of -0.553 and a standard deviation of 0.152.  The values for $\alpha_2$ have a mean of -1.460 and a standard deviation of 0.181 and are also in agreement with the predicted value of -3/2 (fast cooling regime), though the main emission is episode is less consistent with fast cooling compared to the other episodes. The $\beta$ parameter has its distribution centered at a mean of 3.498 and a standard deviation of 0.440.

\begin{figure}[]
    \includegraphics[width=1.0\columnwidth]{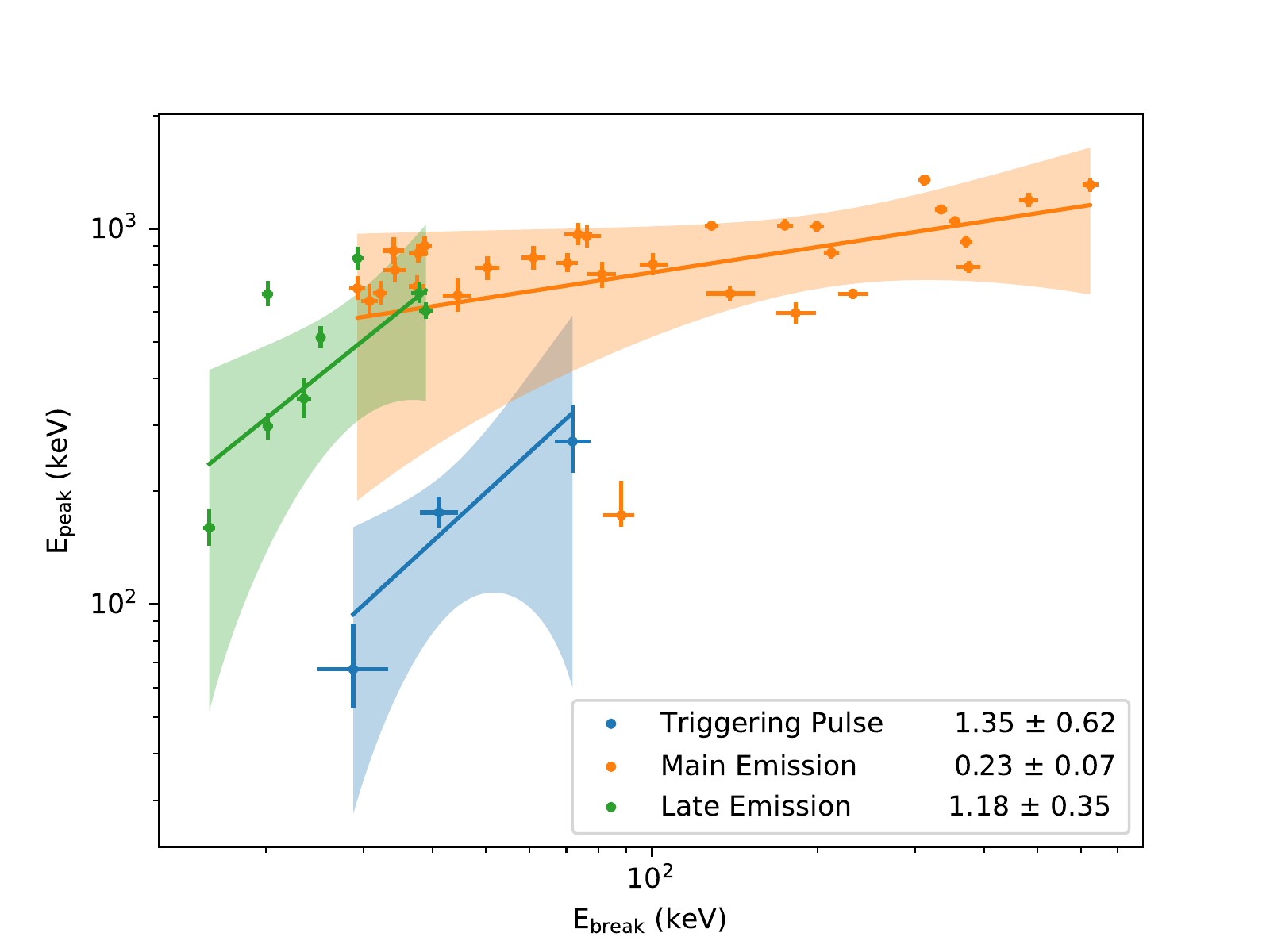}
    \caption{Correlation between peak energy $E_{peak}$ and break energy $E_{break}$ for 2SBPL. The values from the triggering pulse, the main emission, and the late emission are shown. Power-law indices are shown in the legend.}
    \label{fig:eb_ep}
\end{figure}

\begin{figure}[]
    \includegraphics[width=1.0\columnwidth]{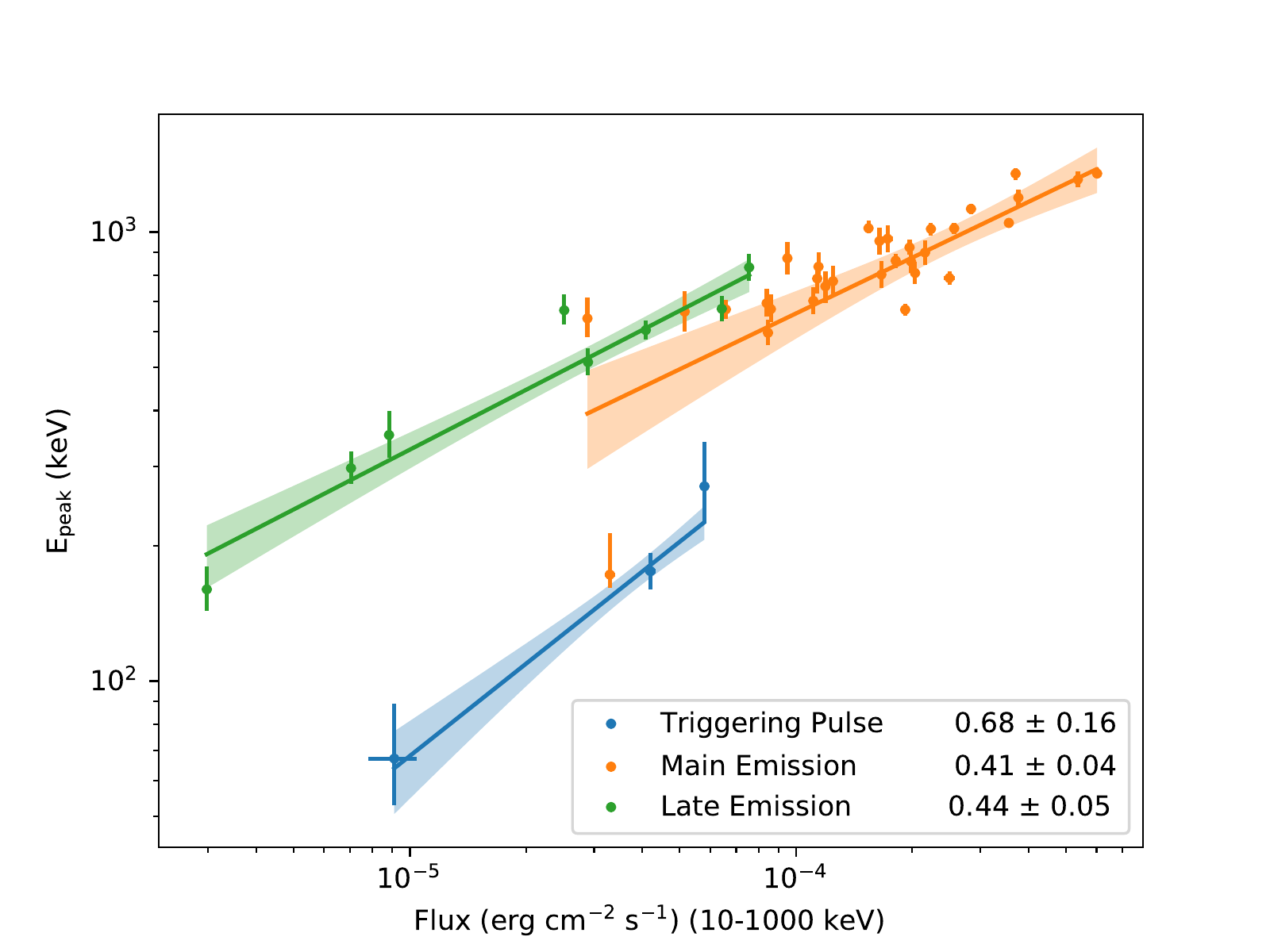}
    \caption{Correlation between peak energy $E_{peak}$ and flux for 2SBPL. The values from the triggering pulse, the main emission, and the late emission are shown. Power-law indices are shown in the legend.}
    \label{fig:ep_flux}
\end{figure}

\begin{figure}[]
    \includegraphics[width=1.0\columnwidth]{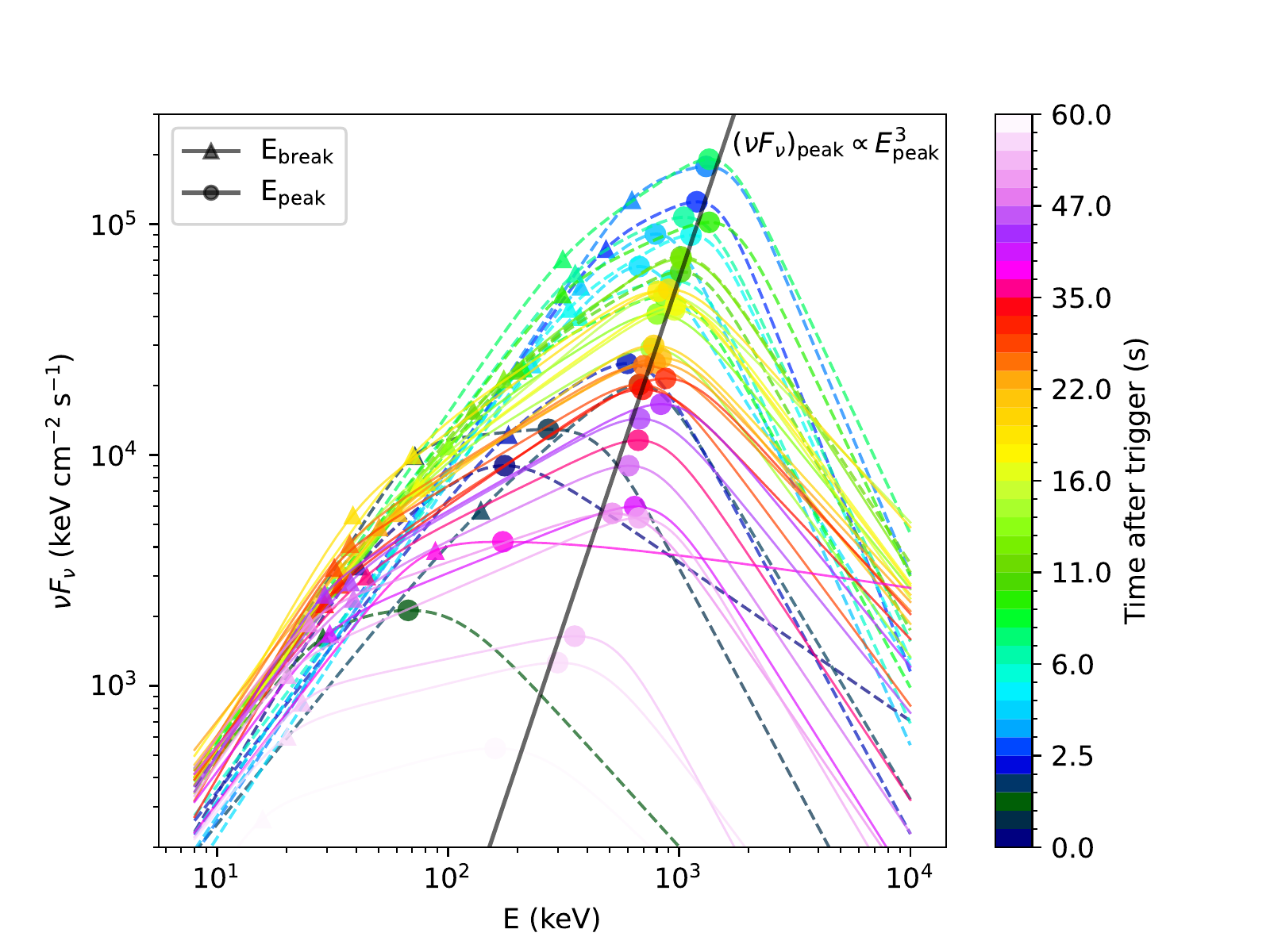}
    \caption{Combined $\nu F_\nu$ spectra for each fitted interval. Colors indicate the time, triangles mark $E_{break}$, circles mark $E_{peak}$. Dashed lines show spectra up to t<10 s, solid lines after. Black line shows the $\nu F_\nu \propto E^3$ relation.}
    \label{fig:nufnupeak}
\end{figure}

The correlation between $E_{peak}$ and $E_{break}$ can be seen in Figure \ref{fig:eb_ep}, where the differing time regions of the triggering pulse, main emission, and secondary emission are highlighted. There is a clear difference in the slopes of the power law fits for the main and late emissions, with the late emission being much steeper. After the dip, the values for $E_{break}$ move below 30~keV. Figure \ref{fig:ep_flux} shows the correlation between $E_{peak}$ and flux, which behave in a similar way for the main and late emission, with the trigger pulse being more distinct. The main and late emission regions have a similar slope of increase in flux, while the triggering pulse shows a much steeper increase. Figure \ref{fig:nufnupeak} shows the temporal evolution of the $\nu F_\nu$ spectra for each fitted interval. Excluding the triggering pulse interval it is found that $\nu F_\nu$ is proportional to $E^3$.

\begin{figure}[t!]
\includegraphics[width=1.0\columnwidth, height=.8\columnwidth]{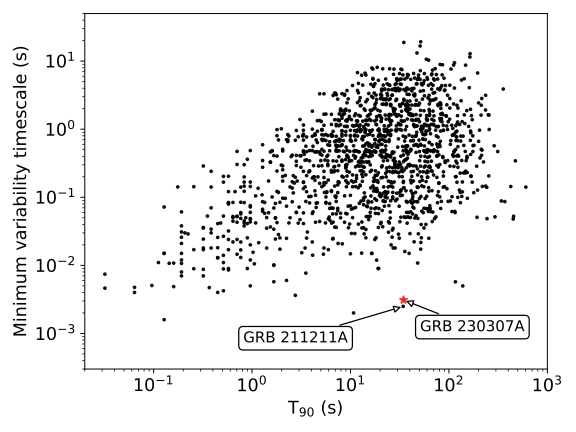}
\caption{MVT values as a compared to T$_{90}$ for all \gbm GRBs with well measured T$_{90}$ and MVT. GRBs 211211A and 230307A, both long bursts with associated kilonovae, appear right near each other and away from the main  distribution. }
\label{fig:manyMVT}
\end{figure}

\subsection{Temporal Properties: Spectral Lag and Minimum Variability Timescale}

The spectral lag measures the time offset between lightcurves in two energy bands: 8-25 and 50-300~keV. In practice, it is found that there is a time offset between the two lightcurves. The spectral lag is potentially an indication of the progenitor of the GRB sGRBs have lags consistent with zero, whereas lGRBs have positive lags \citep{Norris+00lag,Becerra+23_210704a}. With positive lag the harder energy band leads the softer band. Calculating the lag over the entire duration of the burst, it is found to be $-0.0164\pm 0.0196$~s. This value is consistent with zero which is typical of short GRBs. Additionally, the lag is found to be consistent with zero if it is measured in the pre-dip and post-dip time intervals and also in across the different energy ranges. \cite{Wang_2023_broken_alpha} also found spectral lags consistent with zero for three time intervals: $t_0 + 0.2-0.4s$,  $t_0 + 0.4-3.0s$, and $t_0 + 7.0-40s$.

The minimum variability timescale (MVT) represents the shortest timescales in which variations of a GRB lightcurve can be observed. \cite{Veres_2023} analysis of GRB~211211A noted that the MVT emerged as a possible discriminator for long duration GRBs with merger origin. A sample of 10 lGRBs with short MVTs <15~ms were found, though only three bursts GRBs, 090720B, 210410A, and 080807, remained as possible candidates after excluding three known bright bursts from supernovae and four bursts whose lightcurves did not show the three emission episode morphology. GRB~230307A shows remarkable similarities with GRB~211211A in both their lightcurve morphology and MVT values. The MVT for GRB~230307A is $3.1\pm  0.7$~ms, while GRB~211211A has a MVT of $2.6\pm  0.9$~ms \citep{Veres_2023}, which places both bursts at the lower extreme of the MVT distribution for both lGRBs and sGRBS  
(Figure \ref{fig:manyMVT}). This further supports the notion of short MVTs being used as an indirect indication that the GRB is of merger origin.

\subsection{Lorentz Factor}
Given the short variability timescale, a lower limit is first placed on the Lorentz factor. Assuming gamma-rays are emitted through internal shocks \citep{Rees+94unsteady}, the Lorentz factor of the outflow can be constrained by requiring that the internal shocks occur above the photosphere. Thus, the internal shock radius ($R_{\rm IS}\approx2\Gamma^2 c \delta t$) must be larger than the photospheric radius ($R_{\rm ph}\approx \sigma_T L/8\pi m_p c^3 \Gamma^3 $). The limit on the Lorentz factor in this scenario becomes:

\begin{multline}
    \Gamma>\left(\frac{\sigma_T L}{16\pi m_p c^4 \delta t_{\rm var}}\right)^{1/5}= \\ =170 \left(\frac{L_{\rm tot}}{5\times 10^{52} ~{\rm erg} ~{\rm s}^{-1}}\right)^{1/5}  \left(\frac{\delta t_{\rm var}}{3.1 ~{\rm ms}}\right)^{1/5}  
\end{multline}

Note that this limit is only meaningful for high luminosities or very short variability timescales. Here $L_{\rm tot}=L_\gamma/\eta =1.3\times 10^{52} {\rm erg} s^{-1}$
is the total luminosity and $\eta$ is the gamma-ray efficiency, assumed to be 20\%.

\subsection{Interpretation as Fast Cooling Synchrotron}

The $\alpha_1$ and $\alpha_2$ indices are consistent with the expectation from fast cooling synchrotron emission. In this picture, the $\alpha_1\approx -2/3$ below the first break corresponds to the slope of individual electron's synchrotron emission. The $\alpha_2\approx-3/2$ corresponds to the spectrum produced by electrons injected into the emitting region with random Lorentz factor $\gamma_m$ and cooling on a timescale that is short compared with the dynamical timescale of the system. 
The expression of the flux density: $ F_\nu\propto E N_E \propto d\gamma_e/dE\propto E^{-1/2}$ (equivalent to $\alpha_2=-3/2$) where $\gamma_e$ is the electron's random Lorentz factor, E represents the photon energy. $N_E$ is the photon number spectrum and all the fitted indices are in this representation. The last step in the above equation was derived utilizing the expression of the typical synchrotron frequency of electrons with  $\gamma_e$ which is $E(\gamma_e)\propto\gamma_e^2$ \citep{Cohen+97relshock}.

Having established that the spectral indices are consistent with the synchrotron fast cooling scenario, $E_{\rm break}$ represents the cooling frequency ($E_c$) and $E_{\rm peak}$ is the typical or injection frequency ($E_m$) of the synchrotron-emitting electron population. Using these characteristic frequencies of the spectrum throughout the GRB, the physical parameters of the outflow can be constrained.

Following \citet{Kumar+08radius},   equations are inverted for $E_c$, $E_m$, and $F_\nu(E_c)$ and the physical parameters (Lorentz factor $\Gamma$, radius $R$, magnetic field $B$) of the emission region are derived \citep[see also][]{Beniamini+13synchrotron}.
For synchrotron emission,

\begin{equation}
\begin{split}
    E_m\propto B \Gamma \gamma_e^2,  \;  F_{\nu,{\rm pk}}\propto B N \gamma D_L^{-2} \text{, and} \\  
    E_c\propto B^{-3} \Gamma^{-1} \Delta t^{-2} (1+Y)^{-2}
\end{split}
\label{eqn:sync_emisssion}
\end{equation}
where $\Delta t$ is the integration time, N is the number of radiating particles, and $D_L$ is the luminosity distance. 
In these calculations, the Compton parameter, $Y$, defined as the ratio of the photon and magnetic field energy densities, is left as a variable. At high energies, there is no detection of any extra spectral components, indicating the inverse Compton contribution that scales with $Y$ is small, $Y<1$. 
Assuming the power law index of the accelerated electron distribution is $p$ ($dN_e/d\gamma_e\propto \gamma_e^{-p}$) can be calculated from the photon index above $E_{\rm peak}$.  Thus $N_E\propto E^{\beta}\propto E^{-(p+2)/2}$, or simply ${\beta} = -(p+2)/2$.

Substituting the spectral parameters during the brightest part (2.75-10.94 s), and making the conservative assumption that $\Delta t$ duration is twice the width of the bins, an average Lorentz factor of 

\begin{multline}
\Gamma\approx1600 \left(\frac{F_{E_{break}}}{ 86~ mJy}\right)^{1/8} \left(\frac{E_{break}}{330~ keV}\right)^{1/8} \left(\frac{E_{peak}}{1100~ keV}\right)^{3/16} \\ 
 \left(\frac{\Delta t}{1 s}\right)^{-3/8} \left(\frac{t_a}{0.5~s}\right)^{1/4} Y^{-3/16} \left(\frac{1+Y}{2}\right)^{1/4}.
\end{multline}

In the above formula,  representative values for the brightest part of the GRB are included. $t_a$ is the time available for electrons to cool, taken as the temporal bin width \citep{Kumar+08radius}.
The standard deviation of the Lorentz factor (calculated from different values in different time-bins) is $\sigma_\Gamma=260$. Using the variability timescale, this translates to an emission radius of 
\begin{equation}
R\approx2\Gamma^2 c \delta t_{\rm var}=9.4\times 10^{14} \left(\frac{\Gamma}{1600}\right)^2 \left(\frac{\delta t_{\rm var}}{3.1 ~{\rm ms}}\right) \cm 
\label{eqn:emisssion_rad}
\end{equation}
The magnetic field is constrained to $\log_{10}(B/[G])=2.71\pm0.15$ or $B\approx510 $ G. 
\\

The simple fireball dynamics suggests there is a maximal attainable Lorentz factor \citep[see however][for extremely high Lorentz factors]{Ioka+10highLorentz,Meszaros+97poynting}. The jet starts accelerating at a radius $R_0$. The smallest value for this radius can be taken as the innermost stable circular radius of the black hole central engine of mass $M_{BH}$, e.g. $R_0=6M_{BH} G/c^2  = 2.7\times10^6 ~(M_{BH}/3M_{\odot}) \cm$.

The acceleration ceases when the Lorentz factor reaches the saturation value $\eta=L_{tot}/\overset{.}{M}c^2 $ at a radius $R_{\rm sat}=R_0 \eta$, where $L_{tot}$ is the total luminosity, $\overset{.}{M}$ is the mass accretion rate in the jet. The Lorentz factor will be highest if the photosphere occurs approximately at the saturation radius. By equating $R_{\rm sat}=R_{\rm phot}$, it is found that:
\begin{eqnarray}
\Gamma_{\max} &\approx& \left(\frac{L_{tot} \sigma_T}{4\pi m_p c^3 R_0}\right)^{1/4}\\
&\approx& 1700 ~L_{\gamma,52}^{1/4}
\left(\frac{\eta_{\rm eff}}{0.2}\right)^{-1/4}\left(\frac{M_{BH}}{3M_{\odot} }\right)^{-1/4}
\end{eqnarray}
This limit is remarkably close to the derived average of 1600, suggesting an extreme GRB. 
The launching radius $R_0$ can also be associated with the observed variability timescale or $R_0=c \Delta t=9 \times 10^7 (\Delta t/{\rm 3.1 ~ms}) \cm$. This yields a maximum Lorentz factor of $\approx$900, marginally inconsistent with value of 1600. This inconsistency could mean that the variability timescale in the BTI region would be even lower than the 3.1 ms measured outside of the BTI. Alternatively, it could mean that the observed variability is not imprinted on the lightcurve close to the central engine, but it originates further out at the dissipation site.

\subsection{Late Emission and High Latitude Emission}

\begin{figure}[t]
\includegraphics[width=1.0\columnwidth]{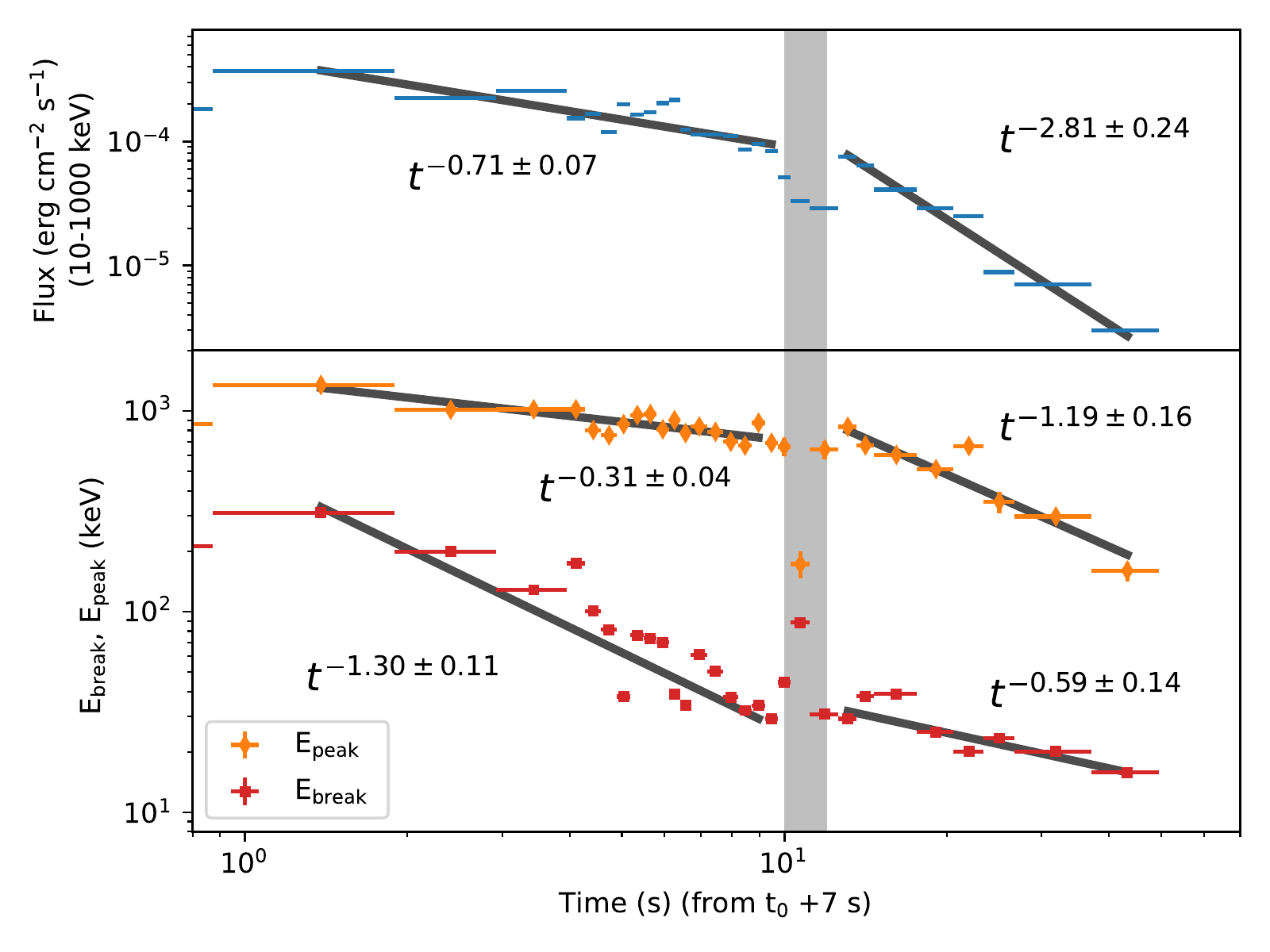}
\caption{The power law indices of $E_{\rm peak}$ (orange) show an initial shallower decline, steepening after the dip, while $E_{\rm break}$ (red) begins steeper and becomes shallower. The steep decline of the flux after the dip is reminiscent of HLE.}
\label{fig:eb_evol}
\end{figure}

The time evolution of $E_{break}$, $E_{peak}$ and the flux show remarkable trends. Their time evolution consists of piecewise power laws (Figure \ref{fig:eb_evol}), and the peak of the $\nu F_\nu$ spectrum is approximately proportional to $E_{peak}^3$  throughout the burst (Figure \ref{fig:nufnupeak}). 

The temporal power law slopes are sensitive to the choice of start time. For GRBs, the start time is typically taken as the trigger time, however in some cases the time of the last significant emission period can be used. Because of the relatively long duration of this burst and prolonged emission episodes, latter approach is chosen and the start time is shifted to $t_0+t_{\rm shift}$ s in Figure \ref{fig:eb_evol} when fitting the time evolution. $t_{\rm shift} =7 s$ is chosen as it corresponds to the last major emission episode (Figure \ref{fig:Fig 1}), noting that this shift is in the middle of the BTI region. 

As noted, the dip (Section \ref{sec:time_resolved}) is another striking feature of this GRB. It is found that it coincides with breaks in the evolution of the  flux, $E_{\rm peak}$ and $E_{\rm break}$ (Figure \ref{fig:eb_evol}). This change in evolution suggests that the dip is not simply a pause in the otherwise continuous string of pulses, but has a physical cause.

Prior to the dip, $E_{\rm peak}$ decays slowly $(t-t_{\rm shift})^{-0.31\pm0.07}$ with a mean of 900~keV. After the dip, $E_{\rm peak}$ decays  at a steeper rate as $(t-t_{\rm shift})^{-1.19\pm0.16}$ (Figure \ref{fig:eb_evol}). $E_{\rm break}$ values vary around 300~keV with no pronounced trend up to $t_{\rm shift}$, where $E_{\rm break}$ starts a steep drop, $(t-t_{\rm shift})^{-1.30\pm 0.11}$ until the dip, then it follows a shallower decay, $(t-t_{\rm shift})^{-0.59\pm0.14}$. The flux, integrated over 10-1000~keV decreases from $t_{\rm shift}$ to the dip as $F\propto(t-t_{\rm shift})^{-0.71\pm 0.07}$, then transitions in to a steeper decay, $F\propto(t-t_{\rm shift})^{-2.81\pm 0.24}$ after the dip.

One of the possible interpretations of the different behavior pre and post-dip is that the dip marks the end of the prompt emission and the start of the afterglow. The temporal power-law indices of $E_{\rm break}$ and $E_{\rm peak}$ evolution are broadly consistent and are within model expectations for afterglow $E_c\propto t^{-1/2}$ and $E_m\propto t^{-3/2}$ respectively \citep[e.g.][]{Sari+98spectra}. However,  the flux decays so fast $\propto (t-t_{\rm shift})^{-2.8}$, that it is impossible to reconcile with the model expectation of $\propto t^{-1/4}$. The afterglow was observed to fade rapidly and be exceedingly faint compared to expectations for a long burst this bright in prompt emission \citep{GCN_33485}.

The steep decay of $E_{\rm break}$ or $E_c$ starts from $t_{\rm shift}$ and lasts until the dip. 
Up to this time, energy was continuously injected into the emission site, $\gamma_c$ remained constant, but as the injection stopped, the electron population responsible for the break in the spectrum started shifting to lower values. 
The cooling break energy corresponds to  a synchrotron-emitting electron, 
that loses its energy (cools) on the dynamic time of the shell, $t'_{dyn} = R/\Gamma c$.
The corresponding synchrotron timescale and frequency are: $t'_{syn}=\gamma_e/(d {\gamma_e}/dt)=3\gamma_e c^2/\sigma_T B^2\gamma_e^2$,  and
$\nu_{\rm syn}(\gamma_e)= {q_e}/{(2\pi m_e c)} B \Gamma \gamma_e^2$  respectively.

Keeping only the relevant variables, the cooling random Lorentz factor can be written $\gamma_c\propto \Gamma B^{-2} R^{-1}$ and utilizing the relations in Equation \eqref{eqn:sync_emisssion} the cooling energy will scale as:
\begin{equation}
    E_c\propto\Gamma B \gamma_c^2\propto \Gamma^3 B^{-3} R^{-2}.
\end{equation}
The injection break will scale as $E_m\propto \Gamma  B \gamma_m^2$. The luminosity of a population of electrons scales as the flux and it is proportional to $L\propto \Gamma^2 B^2 \gamma_m^2$. 

The prompt emission happens in the coasting phase of the jet evolution, where $\Gamma$ is approximately constant. Thus, the emission radius will be proportional to time, $R\propto t$. There are multiple ways to treat the evolution of the magnetic field in the literature e.g. assuming  the flux freezing limit $B\propto R^{-2}$ \citep{dermer04}. 
\citet{Uhm+14Bfield} take a more general approach and parametrize the evolution of the magnetic field as a power law with index $q$: $B\propto R^{-q}$, starting at the emission radius.  
In the simplest model, the injection Lorentz factor ($\gamma_m$) remains constant. This is almost consistent with the observations, as $E_m$ is changing slowly, as $\propto t^{-0.3}$. To allow for this change, the evolution of $\gamma_m \propto t^{-m}$ is parameterized.  
Using these dependencies, it can be found such that $E_c\propto t^{3q-2}$ , $E_m\propto t^{-q-2m}$ and $F\propto t^{-2q-2m}$.
From observations from $t_{\rm shift}=7s$ to the dip, it is derived that $E_c\propto t^{-1.3}$, $E_m\propto t^{-0.3}$ and $F\propto t^{-0.7}$. The solution of this over-determined set of equations is  $m\approx0.04$ and
$q\approx0.29$, which offers a consistent picture of the evolution of the synchrotron parameters.

The steep decline of the flux ($F\propto (t-t_{\rm shift})^{-2.81\pm0.24}$) is reminiscent of observations by Swift XRT  where a steep decline in flux is observed after the end of the prompt emission phase for numerous GRBs \citep{Nousek+06XRTcanonical,Zhang+06ag,Grupe+13plateau}. This is widely attributed to the high latitude emission (HLE) \citep{kumar00, dermer04}. In this scenario, the emission region stops emitting and  delayed emission from progressively larger latitudes of the jet is observed. Considering a power law spectrum with spectral index $\beta_h$ ($F_{\nu}\propto 
 t^{-\alpha_h} \nu^{-\beta_h}$), the HLE predicts a temporal evolution index $\alpha_h=2+\beta_h$. HLE is usually identified in X-rays, in the integrated 0.3-10~keV band flux. Because the spectra of GRB~230307A
is composed of three power law segments and it displays strong spectral evolution, the HLE closure relation is tested for the three spectral regimes, defined by (1)~$E<E_{\rm break}$, (2)~$E_{\rm break}<E<E_{\rm peak}$, and (3)~$E>E_{\rm peak}$ separately. In this notation, the spectral index of the low energy segment of the 2SBPL is $\beta_{h1}=-(\alpha_1+1)$, in the mid- segment it is $\beta_{h2}= -(\alpha_2+1)$ and in the highest energy segment $\beta_{h3}= -(\beta+1)$. 
Thus, the expected temporal decay indices are $\alpha_{h{\{1,2,3\}}}=2+\beta_{h{\{1,2,3\}}}$. 

\begin{table}[]
    \centering
    \begin{tabular}{c|c|c| c}
       &   $E<E_{\rm break}$    & $E_{\rm break}< E <E_{\rm peak}$ & $E_{\rm peak}<E$   \\
       & 1 & 2 & 3 \\
       \hline
    $\alpha_{h}$  & $1.61 \pm 0.13$ & $2.61 \pm 0.13$ & $4.42 \pm 0.26$ \\ \hline
    $\alpha_{meas}$  & $1.12 \pm 0.20$ & $2.66 \pm 0.21$ & $4.44 \pm 0.55$ \\ \hline
    \end{tabular}
    \caption{Table of expected  temporal decay indices in the 2SBPL and the measured temporal indices representative of the three power law segments related to the HLE.}
    \label{tab:spec_temp_ind}
\end{table}

The slopes of the flux density lightcurves for representative energies of the three power law segments, $E_{\{1,2,3\}}=\{20,300, 1000\} $ ~keV, are the measured temporal indices are $\alpha_{\rm meas}$. Table \ref{tab:spec_temp_ind} shows that $\alpha_{\rm meas}$ are in reasonable agreement with $\alpha_{h{\{1,2,3\}}}$, thus it is concluded that the interval after the dip and until the last GBM detection is well described by the high latitude emission.

Assuming that the late emission is due to the high latitude effect, it can used it to constrain the emission radius. For the HLE, $\Delta t_{\rm tail}=R_\gamma \theta^2/2c$, where $\Delta t_{\rm tail}$ is the duration of the HLE emission, $R_\gamma$ is the radius of the gamma-ray emission, and theta is the angle ($\theta>\Gamma^{-1}$) from where the late photons are emitted. 

Taking $\Delta t_{\rm tail}>76 $~s  (19 to 95 s) to include both late emission and the tail, and assuming $\theta=10^{\circ}$ , the radius of emission of gamma-rays can be constrained:   
\begin{equation}
    R_{\gamma}>1.5\times10^{14} \left(\frac{\Delta t_{\rm tail}}{76 ~{\rm s}}\right)  \left(\frac{\theta}{10^{\circ}}\right)^{-2}  {\rm cm}.
\end{equation}
This is consistent with the radius estimate from the synchrotron modeling from Equation \eqref{eqn:emisssion_rad}.

\subsection{In Context of Other GRBs}
\begin{figure}[h]
\includegraphics[width=1.0\columnwidth]{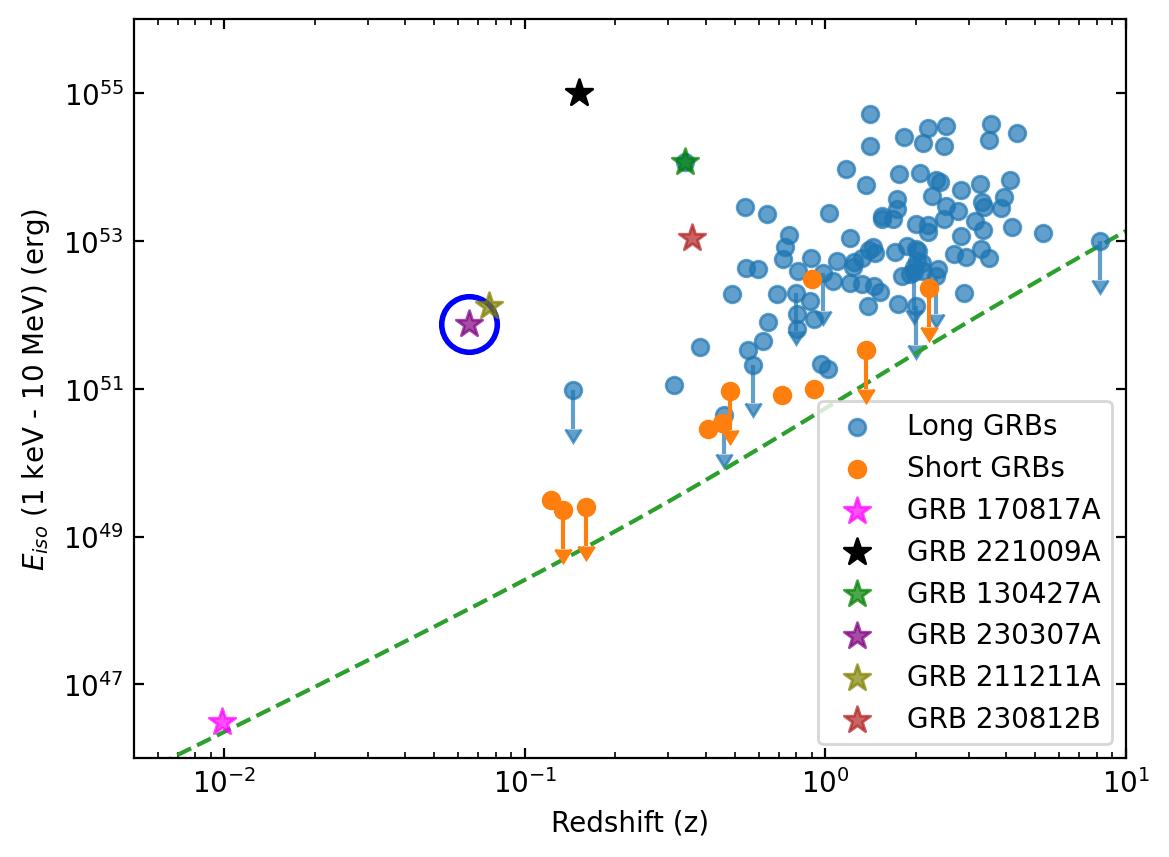}
\includegraphics[width=1.0\columnwidth]{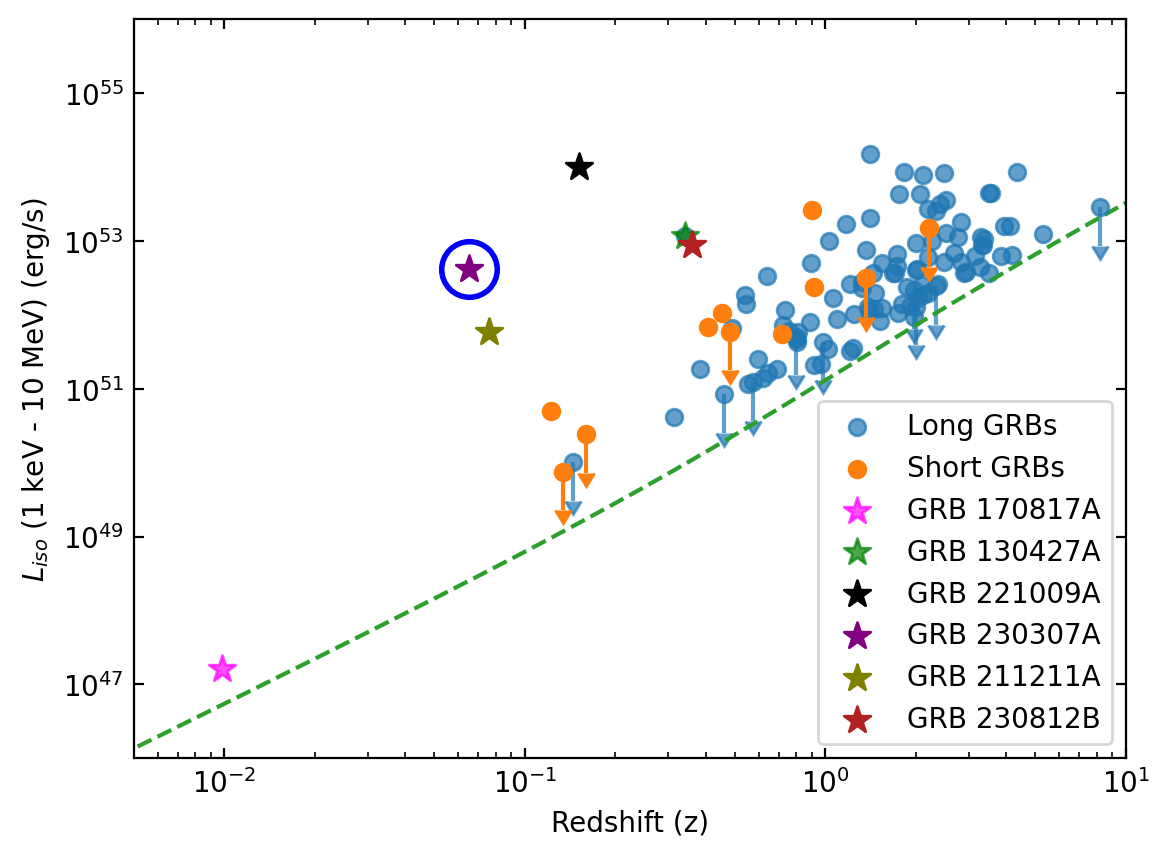}
\caption{Distribution of calculated E$_{iso}$ and L$_{iso}$ values for all \gbm GRBs with well-measured redshifts through 2017 \citep{Abbott_170817A} and updated with measurements from \cite{Poolakkil_GBM_10yr}. Notable GRBs are highlighted.} 
\label{fig:eiso}
\end{figure}
The fluence of \grb was measured to be (6.020 $\pm$ 0.021)$\times$10$^{-3}$ erg cm$^{-2}$ in the 10-10,000~keV band, which makes it only second to GRB 221009A \citep{Burns_2023}. Using the reported redshift of z=0.065 for the host galaxy \citep{GCN_33485}  the total  isotropic-equivalent gamma-ray energy of \grb calculated in the 1-10,000~keV range is $E_{\rm iso} = (6.973\pm 0.016) \times 10^{52}\erg$.  The peak luminosity calculated on the 64~ms timescale is $L_{\rm iso, 64ms} = (1.225 \pm 0.008) \times 10^{52}\erg s^{-1}$. These values for fluence, $E_{\rm iso}$, and $L_{\rm iso}$  are in agreement with those reported by Konus-Wind  \citep{GCN_33427}.  Figure \ref{fig:eiso} places the $E_{\rm iso}$ and  $L_{\rm iso}$  values for \grb within a distribution of \gbm GRBs.  

 The inferred Lorentz factor of $\Gamma= 1600$ for \grb is one of the highest calculated Lorentz factors for any GRBs. \cite{Ghirlanda+18Lorentz} looked at sample of 66 long GRBs and one short with known redshifts and a sample an additional 85 GRBs with known afterglow onset times and found a range of $200 < \Gamma < 700$ with an median value of $\Gamma \sim 300$. \citet{Veres_2023} found $\Gamma \approx 900$ for GRB~211211A. 
Another method for deriving the Lorentz factor is the requirement that high energy (typically GeV range) photons can escape the emission site. The one short GRB was GRB~090510 which was found to have a lower limit at $\Gamma \geq 1200$ \citep{Ackermann+10_GRB090510}. Other large Lorentz factor for GRBs include GRB~090423 with $\Gamma \sim 1100$ \citep{Ruffin_2014}, GRB~080916C and GRB~090902B with $\Gamma = 887$ and $\Gamma = 867$ respectively \citep{Ackermann+12constrain}. 
\cite{Beniamini+13synchrotron} suggest that synchrotron modeling allows for a large range of $\Gamma$  ($300 < \Gamma <3000$). 
The temporal and spectral similarities between \grb and GRB~211211A are numerous:  overall pulse structures,  short MVT, similar T$_{90}$, redshift, close $E_{\rm iso}$, and $L_{\rm iso}$ values. Both bursts are also two of the brightest observed by \gbm and the second and third nearest with confirmed redshifts. It is difficult to say if these similarities are possible traits of this long merger class or coincidence of the two observed GRBs. \cite{Peng_2024} explored a significant number of temporal and spectral properties of both GRBs, including the three emission phase structure, their respective positions along the Amati relation, and the photospheric emissions. Further searches into other possible long mergers, such as in \cite{Veres_2023} have proposed a few possible candidates, but a more in-depth exploration using updated commonalities should be conducted.

\section{Summary} \label{sec:summary}
\grb is the second of the brightest and second closest GRBs ever observed and allows for an unprecedented look into a burst that defies the current GRB duration-based classification scheme. This work reports the unified evolution of \grb with the pulse pile-up corrected data for fine time spectral analysis. Using the 2SBPL model, spectral parameters were found that were consistent with the expected values for synchrotron emission in the fast cooling regime. Additionally, it was noted that the relationships of $E_{peak}$ and $E_{break}$ can be used to constrain the physical parameters of the outflow and result in one of the highest calculated Lorentz factors of $\Gamma= 1600$ for any GRB. The variation in the flux at the later time intervals exhibits characteristics attributed to high latitude emission.  

The evidence of a short MVT of $3.1\pm  0.7$~ms and spectral lags consistent with zero further support the merger interpretation of \grb. While both GRB~211211A and \grb first had their merger origin suggested by later observations of associated kilonovae, they exhibit similarities in MVT, spectral lags, and light curves with extended emission episodes. It can be proposed that these features could be used to distinguish merger-origin GRBs regardless of their duration. It is of note that both of these GRBs are among the brightest and most fluent of \gbm GRBs, and there may be more long-duration GRBs from mergers that have not been identified. The spectral and temporal properties, such as the MVT, spectral lag, and three emission phase structure of \grb suggest the need for a new classification system to better classify between GRBs produced by massive core collapse and those produced by compact binary mergers.

\section{Acknowledgments} \label{sec:acknowledgments}

The UAH coauthors gratefully acknowledge NASA funding from cooperative agreement 80MSFC22M0004. The USRA coauthors gratefully acknowledge NASA funding from cooperative agreement 80NSSC24M0035. C.M. is supported by INAF (Research Grant `Uncovering the optical beat of the fastest magnetised neutron stars 620 (FANS)') and the Italian Ministry of University and Research (MUR) (PRIN 2020, Grant 2020BRP57Z, `Gravitational and Electromagnetic-wave Sources in the Universe with current and next-generation detectors (GEMS)').

\bibliographystyle{aasjournal}
\bibliography{references}
\begin{deluxetable}{ccccccccc}[h!]
    \label{tab:params}
 \tabletypesize{\footnotesize}
 \tablecolumns{7}
 \tablecaption{Double Smoothly Broken Power Law Fitting }
 \tablehead{   
   \colhead{Time (s)} &
   {$E_{peak}$} &
   {$E_{break}$} &
   {$\alpha_{1}$} &
   {$\alpha_{2}$} &
   {$\beta$} & 
   {$A$} &
   {$P_{stat}$/DoF} 
    }
\vspace{-0.3cm}
\startdata
 -0.064-0.128 & $175.62^{+17.55}_{-15.83}$ & $41.14^{+3.36}_{-3.14}$ & $-0.301^{+0.016}_{-0.017}$ & $-1.202^{+0.076}_{-0.079}$ & $-2.687^{+0.231}_{-0.257}$ & $6.805^{+0.418}_{-0.404}$ & $70/225$ \\
 0.128-0.202 & $271.29^{+69.15}_{-47.60}$ & $71.88^{+5.62}_{-5.20}$ & $-0.436^{+0.019}_{-0.020}$ & $-1.884^{+0.168}_{-0.099}$ & $-3.867^{+0.644}_{-0.688}$ & $14.149^{+1.143}_{-1.083}$ & $64/225$ \\
 0.202-0.355 & $67.19^{+21.87}_{-14.32}$ & $28.75^{+4.53}_{-4.04}$ & $-0.738^{+0.034}_{-0.034}$ & $-1.751^{+0.355}_{-0.222}$ & $-3.108^{+0.494}_{-0.580}$ & $26.799^{+3.094}_{-2.728}$ & $71/225$ \\
 0.355-1.064 & $671.91^{+33.36}_{-32.16}$ & $138.63^{+15.00}_{-13.14}$ & $-0.758^{+0.007}_{-0.007}$ & $-1.194^{+0.040}_{-0.041}$ & $-3.688^{+0.304}_{-0.337}$ & $14.410^{+0.455}_{-0.455}$ & $93/225$ \\
 1.064-1.404 & $595.99^{+40.62}_{-37.05}$ & $182.32^{+16.17}_{-14.19}$ & $-0.726^{+0.008}_{-0.008}$ & $-1.424^{+0.076}_{-0.076}$ & $-3.854^{+0.359}_{-0.445}$ & $18.466^{+0.715}_{-0.699}$ & $81/225$ \\
 1.404-1.726 & $1191.28^{+52.28}_{-49.18}$ & $482.36^{+18.95}_{-18.21}$ & $-0.542^{+0.004}_{-0.004}$ & $-1.560^{+0.080}_{-0.083}$ & $-4.536^{+0.262}_{-0.280}$ & $10.906^{+0.264}_{-0.256}$ & $101/225$ \\
 1.726-2.034 & $1307.51^{+57.56}_{-53.04}$ & $624.18^{+20.83}_{-20.17}$ & $-0.469^{+0.004}_{-0.004}$ & $-1.657^{+0.108}_{-0.109}$ & $-4.380^{+0.215}_{-0.224}$ & $7.653^{+0.171}_{-0.167}$ & $93/225$ \\
 2.034-2.347 & $789.23^{+29.08}_{-28.49}$ & $375.18^{+19.40}_{-18.76}$ & $-0.504^{+0.005}_{-0.005}$ & $-1.338^{+0.091}_{-0.123}$ & $-4.225^{+0.241}_{-0.261}$ & $8.686^{+0.235}_{-0.232}$ & $111/225$ \\
 2.347-2.752 & $671.00^{+21.22}_{-20.46}$ & $231.56^{+15.20}_{-13.96}$ & $-0.454^{+0.005}_{-0.005}$ & $-1.040^{+0.049}_{-0.057}$ & $-3.611^{+0.196}_{-0.201}$ & $6.246^{+0.168}_{-0.163}$ & $95/225$ \\
  \hline & & & BTI \\ \hline
2.752-3.776 & $1124.73^{+30.65}_{-29.90}$ & $334.71^{+8.57}_{-8.14}$ & $-0.563^{+0.003}_{-0.003}$ & $-1.434^{+0.030}_{-0.030}$ & $-4.209^{+0.154}_{-0.154}$ & $11.601^{+0.169}_{-0.175}$ & $162/225$ \\
 3.776-4.800 & $923.31^{+35.76}_{-33.53}$ & $371.32^{+9.99}_{-9.70}$ & $-0.669^{+0.003}_{-0.003}$ & $-1.690^{+0.054}_{-0.063}$ & $-4.181^{+0.152}_{-0.180}$ & $17.272^{+0.285}_{-0.289}$ & $127/225$ \\
 4.800-5.824 & $1046.74^{+27.08}_{-26.30}$ & $354.72^{+8.08}_{-8.13}$ & $-0.655^{+0.002}_{-0.002}$ & $-1.535^{+0.033}_{-0.033}$ & $-4.160^{+0.125}_{-0.133}$ & $25.992^{+0.335}_{-0.338}$ & $161/225$ \\
 5.824-6.848 & $1348.00^{+27.96}_{-24.77}$ & $313.61^{+6.51}_{-5.65}$ & $-0.542^{+0.002}_{-0.002}$ & $-1.339^{+0.019}_{-0.016}$ & $-4.131^{+0.106}_{-0.116}$ & $18.538^{+0.205}_{-0.205}$ & $251/225$ \\
 6.848-7.872 & $862.50^{+31.38}_{-30.96}$ & $211.74^{+6.84}_{-6.36}$ & $-0.739^{+0.003}_{-0.003}$ & $-1.520^{+0.029}_{-0.027}$ & $-3.833^{+0.131}_{-0.205}$ & $31.705^{+0.517}_{-0.522}$ & $111/225$ \\
 7.872-8.896 & $1347.62^{+39.87}_{-43.67}$ & $311.43^{+6.61}_{-7.16}$ & $-0.671^{+0.002}_{-0.003}$ & $-1.551^{+0.021}_{-0.025}$ & $-4.082^{+0.155}_{-0.131}$ & $27.544^{+0.361}_{-0.366}$ & $159/225$ \\
 8.896-9.920 & $1014.56^{+32.65}_{-33.22}$ & $199.08^{+6.25}_{-6.09}$ & $-0.703^{+0.003}_{-0.003}$ & $-1.411^{+0.022}_{-0.021}$ & $-3.789^{+0.140}_{-0.158}$ & $27.748^{+0.428}_{-0.436}$ & $126/225$ \\
 9.920-10.944 & $1018.06^{+29.24}_{-28.32}$ & $128.36^{+3.65}_{-3.56}$ & $-0.579^{+0.003}_{-0.003}$ & $-1.252^{+0.013}_{-0.014}$ & $-3.520^{+0.121}_{-0.136}$ & $17.980^{+0.276}_{-0.264}$ & $154/225$ \\
  \hline & & &  \\ \hline
10.944-11.274 & $1018.80^{+42.26}_{-24.68}$ & $174.30^{+5.91}_{-5.52}$ & $-0.668^{+0.004}_{-0.003}$ & $-1.319^{+0.020}_{-0.018}$ & $-3.561^{+0.322}_{-0.058}$ & $25.257^{+0.465}_{-0.341}$ & $177/234$ \\
 11.274-11.577 & $803.09^{+56.49}_{-53.47}$ & $100.53^{+6.41}_{-5.55}$ & $-0.676^{+0.007}_{-0.007}$ & $-1.354^{+0.030}_{-0.029}$ & $-3.247^{+0.270}_{-0.284}$ & $27.843^{+0.879}_{-0.864}$ & $87/225$ \\
 11.577-11.889 & $756.20^{+60.37}_{-60.98}$ & $81.21^{+4.93}_{-4.79}$ & $-0.669^{+0.008}_{-0.008}$ & $-1.401^{+0.030}_{-0.031}$ & $-3.363^{+0.325}_{-0.358}$ & $25.093^{+0.885}_{-0.855}$ & $100/225$ \\
 11.889-12.185 & $859.29^{+52.00}_{-48.25}$ & $37.74^{+1.56}_{-1.42}$ & $-0.330^{+0.008}_{-0.009}$ & $-1.167^{+0.015}_{-0.016}$ & $-3.050^{+0.244}_{-0.257}$ & $10.626^{+0.326}_{-0.321}$ & $101/225$ \\
 12.185-12.494 & $953.90^{+68.99}_{-62.92}$ & $76.23^{+4.57}_{-4.30}$ & $-0.635^{+0.008}_{-0.008}$ & $-1.303^{+0.023}_{-0.023}$ & $-3.398^{+0.311}_{-0.337}$ & $22.936^{+0.749}_{-0.722}$ & $102/225$ \\
 12.494-12.799 & $965.19^{+68.14}_{-63.56}$ & $73.49^{+4.39}_{-4.17}$ & $-0.627^{+0.007}_{-0.008}$ & $-1.283^{+0.022}_{-0.022}$ & $-3.357^{+0.297}_{-0.324}$ & $22.648^{+0.723}_{-0.717}$ & $104/225$ \\
 12.799-13.108 & $809.61^{+49.57}_{-44.73}$ & $70.34^{+3.07}_{-3.08}$ & $-0.560^{+0.006}_{-0.007}$ & $-1.322^{+0.020}_{-0.021}$ & $-3.329^{+0.240}_{-0.252}$ & $24.934^{+0.686}_{-0.699}$ & $83/225$ \\
 13.108-13.409 & $899.29^{+55.95}_{-55.27}$ & $38.75^{+1.21}_{-1.19}$ & $-0.264^{+0.008}_{-0.008}$ & $-1.266^{+0.015}_{-0.015}$ & $-3.129^{+0.230}_{-0.238}$ & $10.962^{+0.313}_{-0.307}$ & $104/225$ \\
 13.409-13.711 & $776.29^{+62.83}_{-55.98}$ & $34.23^{+1.63}_{-1.58}$ & $-0.433^{+0.010}_{-0.010}$ & $-1.265^{+0.019}_{-0.019}$ & $-3.106^{+0.300}_{-0.346}$ & $14.359^{+0.519}_{-0.499}$ & $111/225$ \\
 13.711-14.201 & $836.27^{+64.50}_{-58.44}$ & $61.00^{+3.05}_{-2.93}$ & $-0.698^{+0.007}_{-0.007}$ & $-1.405^{+0.021}_{-0.021}$ & $-3.237^{+0.265}_{-0.290}$ & $30.436^{+0.881}_{-0.854}$ & $70/225$ \\
 14.201-14.708 & $786.94^{+57.96}_{-57.59}$ & $50.32^{+2.58}_{-2.39}$ & $-0.720^{+0.007}_{-0.007}$ & $-1.392^{+0.017}_{-0.022}$ & $-3.088^{+0.236}_{-0.268}$ & $36.938^{+1.052}_{-1.004}$ & $96/225$ \\
 14.708-15.209 & $702.08^{+50.84}_{-46.96}$ & $37.55^{+1.28}_{-1.25}$ & $-0.453^{+0.008}_{-0.008}$ & $-1.379^{+0.017}_{-0.017}$ & $-3.061^{+0.228}_{-0.247}$ & $17.265^{+0.484}_{-0.471}$ & $97/225$ \\
 15.209-15.707 & $672.99^{+52.81}_{-44.27}$ & $32.19^{+0.91}_{-0.95}$ & $-0.236^{+0.009}_{-0.009}$ & $-1.397^{+0.017}_{-0.018}$ & $-3.349^{+0.298}_{-0.319}$ & $8.128^{+0.253}_{-0.238}$ & $105/225$ \\
 15.707-16.211 & $873.28^{+76.53}_{-68.25}$ & $34.04^{+1.53}_{-1.48}$ & $-0.565^{+0.009}_{-0.009}$ & $-1.355^{+0.017}_{-0.017}$ & $-3.117^{+0.288}_{-0.318}$ & $19.900^{+0.626}_{-0.613}$ & $101/225$ \\
 16.211-16.728 & $693.88^{+52.19}_{-47.03}$ & $29.25^{+0.99}_{-0.96}$ & $-0.259^{+0.009}_{-0.010}$ & $-1.306^{+0.017}_{-0.017}$ & $-3.064^{+0.260}_{-0.296}$ & $7.247^{+0.234}_{-0.226}$ & $107/225$ \\
 16.728-17.264 & $664.19^{+74.02}_{-63.45}$ & $44.48^{+2.67}_{-2.62}$ & $-0.773^{+0.010}_{-0.010}$ & $-1.508^{+0.026}_{-0.027}$ & $-3.516^{+0.445}_{-0.520}$ & $32.428^{+1.171}_{-1.160}$ & $97/225$ \\
 17.264-17.849 & $573.06^{+91.31}_{-46.69}$ & $33.41^{+1.04}_{-1.88}$ & $-0.442^{+0.009}_{-0.014}$ & $-1.512^{+0.025}_{-0.027}$ & $-3.376^{+0.558}_{-0.390}$ & $10.021^{+0.341}_{-0.441}$ & $93/225$ \\
 17.849-18.434 & $223.09^{+70.97}_{-50.22}$ & $19.73^{+1.36}_{-1.43}$ & $-0.512^{+0.023}_{-0.025}$ & $-1.691^{+0.058}_{-0.062}$ & $-3.090^{+0.787}_{-0.857}$ & $9.198^{+0.648}_{-0.651}$ & $69/225$ \\
 18.434-19.019 & $449.00^{+146.45}_{-89.39}$ & $23.54^{+1.27}_{-1.26}$ & $-0.417^{+0.017}_{-0.018}$ & $-1.642^{+0.039}_{-0.039}$ & $-2.895^{+0.597}_{-0.845}$ & $7.850^{+0.448}_{-0.423}$ & $103/225$ \\
 19.019-19.604 & $588.58^{+68.55}_{-50.50}$ & $27.68^{+0.85}_{-0.84}$ & $-0.263^{+0.010}_{-0.011}$ & $-1.504^{+0.020}_{-0.021}$ & $-3.088^{+0.279}_{-0.434}$ & $8.718^{+0.288}_{-0.290}$ & $97/225$ \\
 19.604-20.623 & $833.50^{+59.84}_{-55.13}$ & $29.26^{+0.74}_{-0.72}$ & $-0.381^{+0.007}_{-0.007}$ & $-1.430^{+0.013}_{-0.013}$ & $-3.226^{+0.234}_{-0.261}$ & $12.024^{+0.286}_{-0.275}$ & $123/225$ \\
 20.623-21.652 & $673.96^{+45.30}_{-41.29}$ & $37.87^{+1.30}_{-1.28}$ & $-0.606^{+0.007}_{-0.007}$ & $-1.431^{+0.016}_{-0.016}$ & $-3.248^{+0.238}_{-0.262}$ & $20.304^{+0.496}_{-0.488}$ & $108/225$ \\
 21.652-24.619 & $604.25^{+31.13}_{-28.29}$ & $38.94^{+1.11}_{-1.09}$ & $-0.813^{+0.005}_{-0.005}$ & $-1.527^{+0.012}_{-0.012}$ & $-3.495^{+0.196}_{-0.217}$ & $35.373^{+0.588}_{-0.588}$ & $117/225$ \\
 24.619-27.595 & $512.81^{+36.73}_{-32.69}$ & $25.10^{+0.53}_{-0.52}$ & $-0.625^{+0.006}_{-0.006}$ & $-1.654^{+0.012}_{-0.012}$ & $-3.468^{+0.244}_{-0.275}$ & $25.388^{+0.460}_{-0.454}$ & $115/225$ \\
 27.595-30.379 & $668.93^{+55.60}_{-47.87}$ & $20.11^{+0.48}_{-0.47}$ & $-0.548^{+0.007}_{-0.007}$ & $-1.567^{+0.012}_{-0.012}$ & $-3.694^{+0.344}_{-0.441}$ & $16.353^{+0.352}_{-0.343}$ & $115/225$ \\
 30.379-33.696 & $352.95^{+46.23}_{-39.15}$ & $23.41^{+0.72}_{-0.70}$ & $-0.629^{+0.009}_{-0.009}$ & $-1.790^{+0.024}_{-0.024}$ & $-3.855^{+0.506}_{-0.549}$ & $12.808^{+0.370}_{-0.357}$ & $120/225$ \\
 33.696-44.043 & $297.78^{+26.16}_{-22.91}$ & $20.12^{+0.45}_{-0.44}$ & $-0.747^{+0.006}_{-0.006}$ & $-1.753^{+0.015}_{-0.015}$ & $-3.358^{+0.258}_{-0.276}$ & $16.061^{+0.301}_{-0.289}$ & $161/225$ \\
 44.043-56.422 & $159.96^{+19.66}_{-16.75}$ & $15.77^{+0.39}_{-0.38}$ & $-0.495^{+0.010}_{-0.010}$ & $-1.711^{+0.023}_{-0.022}$ & $-2.990^{+0.248}_{-0.276}$ & $4.758^{+0.128}_{-0.123}$ & $137/225$ \\
56.422-95.770 & $27.16^{+0.81}_{-0.71}$ & $35.22^{+964.78}_{-6.23}$ & $-0.329^{+0.011}_{-0.011}$ & $-1.740^{+0.010}_{-0.010}$ & $-2.235^{+0.011}_{-0.009}$ & $0.832^{+0.031}_{-0.025}$ & $286/225$ \\
\enddata
\vspace{-0.5cm}
\end{deluxetable}


\end{document}